\documentclass{article}

\usepackage{arxiv}

\usepackage[utf8]{inputenc} 
\usepackage[T1]{fontenc}    
\usepackage{hyperref}       
\usepackage{url}            
\usepackage{booktabs}       
\usepackage{amsfonts}       
\usepackage{nicefrac}       
\usepackage{microtype}      
\usepackage{lipsum}
\usepackage{multicol}
\usepackage{setspace}
\usepackage[numbers]{natbib}
\usepackage{subcaption}
\usepackage{graphicx}
\usepackage[table]{xcolor}
\usepackage{amsmath}
\usepackage{amssymb}
\usepackage{algorithm}
\usepackage{enumitem}   
\usepackage{mathtools}
\usepackage{amsthm}
\usepackage{multirow}
\usepackage{algpseudocode}
\usepackage{physics}  
\graphicspath{ {./images/} }
\newcommand*{\ngradpar}{\nabla_{\parallel}}

\newcommand*{\vpare}{V_{\parallel e}}
\newcommand*{\tempe}{T_{ e}}
\newcommand*{\tempi}{T_{ i}}
\newcommand*{\vpari}{V_{\parallel i}}
\newcommand*{\ngradperp}{\nabla_{\perp}}
\newcommand*{\nlaplpar}{\nabla_{\parallel}^2}
\newcommand*{\nlaplperp}{\nabla_{\perp}^2}
\definecolor{darkgreen}{RGB}{0,143,18}

\algblock{Inner}{EndInner}
\algtext*{Inner}
\algtext*{EndInner}

\title{Implicit-Explicit time integration scheme with Physics-based preconditioning for two-fluid tokamak boundary simulations}

\author{
  Micol Bassanini$^{1,2}$,
  Simone Deparis$^{2}$,
  Paolo Ricci$^{1}$,
  Brenno De Lucca$^{1}$,\\
  \textbf{Davide Mancini$^{1}$,
  Louis Stenger$^{1}$,
  Sergio García Herreros$^{1}$
   }\\
  \\
  $^{1}$ École polytechnique fédérale de Lausanne (EPFL), Swiss Plasma Center (SPC), Lausanne, 1015, Switzerland  \\
  $^{2}$ École polytechnique fédérale de Lausanne (EPFL), Institute of Mathematics, Lausanne, 1015, Switzerland \\
}

\begin{document}
\maketitle
\bigskip\bigskip
\begin{abstract}


In this work, a globally stiffly accurate Implicit–Explicit (IMEX) Runge–Kutta scheme
is developed and implemented in the GBS code [Ricci et al., \textit{Plasma Phys. Control. Fusion}, 2012], for two-fluid plasma turbulence simulations. The stiffest phenomena, governed by shear Alfvén waves and parallel diffusion, are treated implicitly, while the remaining non-stiff terms are advanced explicitly. This splitting enables time steps well beyond the Courant–Friedrichs–Lewy limit, without incurring the full computational cost of a globally implicit formulation. To efficiently solve the implicit subsystem at each time step, a three-dimensional physics-based preconditioner, inspired by techniques developed in the magnetohydrodynamic (MHD) context, is used. 
The resulting framework is verified through the method of manufactured solutions and exhibits both algorithmic and parallel scalability.
Significant advantages in numerical stability and computational efficiency are demonstrated with respect to an adaptive explicit Runge-Kutta scheme. 


\end{abstract}


\section{Introduction}
\label{sec:intro}

The boundary region of tokamaks plays a key role in regulating both core plasma behavior and plasma-surface interactions, and therefore strongly influences overall device performance. Reliable modeling of the boundary is essential not only for interpreting experimental observations and improving our theoretical understanding, but also for guiding the operation and design of future fusion reactors such as ITER through the development of predictive scaling laws \cite{giacomin2022first, lim2024predictive}.
From a numerical standpoint, simulating the tokamak boundary is particularly demanding. The difficulty stems from the need to simultaneously resolve turbulent phenomena spanning a wide range of spatial and temporal scales, from the fast, small-scale fluctuations to the slower, large-scale transport processes. Moreover, because of the strong magnetic field present in these devices, particle motion is constrained across magnetic field lines, while it is not along them, resulting in strongly anisotropic dynamics. Accurately capturing this multi-scale, anisotropic turbulent behavior requires both sophisticated physical models and highly efficient computational methods.

Turbulent phenomena in the plasma boundary are commonly investigated using models based on the drift-reduced Braginskii equations~\cite{braginskii1965transport}, implemented in several state-of-the-art codes such as GBS~\cite{giacomin2022gbs}, GDB~\cite{zhu2018gdb}, GRILLIX~\cite{stegmeir2025grillix}, Hermes-3~\cite{dudson2024hermes} (based on the BOUT++ library~\cite{dudson2015bout++}), HESEL~\cite{thrysoe2018plasma}, and SOLEDGE3X~\cite{sureshkumar2024first}. A central numerical challenge in these codes is their efficient time integration, due to the broad range of temporal scales. 

Several strategies have been developed to integrate the drift-reduced Braginskii equations in time, which can be broadly grouped into three families according to their treatment of the stiff, fast dynamics. 
A first approach uses an explicit time integration but subcycles the fast phenomena with a smaller time step, as in GDB, where an explicit second-order trapezoidal leapfrog scheme is used, with subcycling employed to handle the stiffness introduced by the thermal conductivity terms. 
A second approach is to adopt a fully implicit time integration and to exploit the structure of the fast dynamics to build an efficient preconditioner. This is done in BOUT++, where a backward differentiation formula (BDF) scheme is combined with Jacobian-free Newton–Krylov (JFNK) and nonlinear GMRES (NGMRES) methods, preconditioned through a physics-based strategy in which the full Jacobian of the system is approximately inverted \cite{dudson2012improved}.
The last approach is to split the dynamics into a stiff part, treated implicitly, and a non-stiff part, treated explicitly. Advanced time integration schemes of this family have been developed for modeling, for example, solar atmosphere, combining efficient explicit stabilized integration for diffusion terms with implicit integration technique for the reaction term, with variable time-stepping \cite{abdulle2013pirock,wargnier2025time}. In the context of the boundary simulations, the splitting strategy is adopted in GRILLIX, where the Karniadakis multi-step method \cite{karniadakis1991high} is employed and the terms related to the parallel current are treated implicitly \cite{stegmeir2018grillix}, and in SOLEDGE3X, where a variable-stepsize implicit-explicit scheme (VSIMEX) \cite{wang2008variable} treats implicitly the fastest processes (ionization and recombination, the resistive and viscous effects of the Spitzer–Härm model, and electron inertia) to avoid overly stringent time-step constraints \cite{dull2025electromagnetic}.

The work presented here is based on the GBS code \cite{ricci2012simulation,halpern2016gbs}, a first-principles, three-dimensional, flux-driven, turbulence code that evolves the drift-reduced Braginskii equations and has been extensively employed to study plasma turbulence in the boundary region of magnetic fusion confinement devices. In its current formulation, GBS performs time integration using an explicit adaptive Runge–Kutta scheme, with all plasma quantities evaluated on a uniform Cartesian grid expressed in cylindrical coordinates.
Here, we adopt a splitting strategy in GBS, treating the stiff terms implicitly while the remaining non-stiff terms are advanced explicitly. For this purpose, we use an Implicit-Explicit Runge–Kutta (IMEX-RK)~\cite{kennedy2003additive} time stepping algorithm.

IMEX-RK methods have been widely adopted by the computational physics and engineering communities for the efficient treatment of multiscale problems governed by both stiff and non-stiff processes.
In the present context, the fastest, stiffest dynamics (associated with shear Alfvén waves, electron and ion heat conduction along magnetic field lines, and electron and ion viscosity) are treated implicitly, thereby removing the severe stability constraints they impose on the time step. The remaining, non-stiff terms are advanced explicitly, preserving computational efficiency and parallel scalability.
This splitting is physically motivated: the timescales associated with the stiffest phenomena are significantly shorter than those of the turbulent fluctuations of primary interest, and their accurate resolution is therefore not required \cite{durran1998numerical}. 
Indeed, the time step dictated by the stiff terms is several orders of magnitude smaller than the time step required for accuracy purposes alone, making a fully explicit treatment prohibitively expensive.
As a result, the IMEX approach allows significantly larger time steps than fully explicit methods which are subject to a Courant–Friedrichs–Lewy (CFL) stability constraint \cite{courant1928partiellen}, while avoiding the prohibitive computational cost and poor scalability associated with inverting the full Jacobian, as required by a fully implicit formulation \cite{mousseau2000physics}.

The overall efficiency of the IMEX approach critically depends on the ability to solve the implicit subsystem effectively, and an effective preconditioning strategy is essential to improve solver performance. 
While algebraic approaches, such as incomplete LU (ILU) factorizations or multigrid methods, treat the system as a purely mathematical construct without invoking any knowledge of the underlying physics, physics-based preconditioners, by contrast, exploit the structure of the governing equations \cite{mousseau2000physics}. Physics-based preconditioners use techniques such as Picard linearization or operator splitting, to construct an approximate inverse of the system that captures the dominant physical processes without requiring the assembly or factorization of the full Jacobian matrix. This problem-specific strategy is particularly well suited to multi-physics systems, as it enables the natural separation of distinct physical processes and timescales, significantly accelerating the convergence of Krylov subspace methods such as GMRES \cite{saad1986gmres} while preserving the accuracy of both transient and steady-state solutions \cite{chacon2002implicit}. Physics-based preconditioning has indeed been previously explored in plasma boundary turbulence codes: in BOUT++, such a preconditioner is coupled to a BDF/JFNK solver and made tractable by exploiting the field-aligned coordinate system, which reduces the three-dimensional problem to a set of independent one-dimensional solves along field lines \cite{dudson2012improved}.

The present work extends the preconditioning approach used in BOUT++ in several respects. Inspired by the physics-based preconditioning strategy developed by Chacón in the MHD context \cite{chacon2002implicit}, we implement a physics-based preconditioner for drift-reduced Braginskii turbulence simulations and combine it, for the first time, with an IMEX-RK time integration scheme, which
avoids the need to approximate the full Jacobian. Furthermore, since the GBS computational grid is not field-aligned, the dimensional reduction, exploited for example in BOUT++ \cite{dudson2012improved}, is not available here; instead, the preconditioner operates directly on the full three-dimensional implicit operator. This three-dimensional treatment remains computationally tractable precisely because the subsystem evolved implicitly is kept minimal. To the best of our knowledge, this is the first coupled IMEX-RK and physics-based preconditioner framework applied to drift-reduced Braginskii turbulence simulations, and we systematically demonstrate its efficiency advantage over fully explicit time integration.

The implementation is carried out using the Portable, Extensible Toolkit for Scientific Computing (PETSc) library \cite{petsc-web-page}, ensuring scalability on modern high-performance computing architectures. The correctness of the implementation is verified through the method of manufactured solutions (MMS) \cite{riva2014verification} combined with order-of-accuracy convergence tests. The computational performance and scalability of the resulting framework are then systematically assessed across multiple grid sizes and compared against those of the explicit time integration scheme Bogacki-Shampine (ODE23) \cite{bogacki19893,shampine1997matlab}. These tests demonstrate that the IMEX approach, combined with the physics-based preconditioning strategy, exhibits favorable scalability properties. Beyond these computational advantages, the implicit treatment of stiff terms allows for the temperature dependence of the thermal conductivity coefficients, which substantially increases the stiffness of the parallel conduction terms, neglected in the original GBS formulation. This represents a significant extension of the physical fidelity of the model, since heat conduction is known to play a fundamental role in the boundary region of the plasma \cite{stangeby2000plasma}.

 This paper is organized as follows. Section~\ref{sec:two-fluid} presents the drift-reduced two-fluid Braginskii equations and the discretization employed in GBS for boundary plasma simulations. Section~\ref{sec:imex} introduces the IMEX-RK time integration framework and details its construction, while Section~\ref{sec:pb_prec} presents the three-dimensional physics-based preconditioner and its implementation in GBS. Numerical results are reported in Section~\ref{sec:num}, where the code is first verified through MMS convergence tests and then subjected to a systematic assessment of scalability and computational efficiency against the explicit time integration scheme. The numerical results include an application of the IMEX-RK method to an experimental magnetic equilibrium, in which the computational efficiency of the IMEX strategy is further benchmarked against the explicit approach. Finally, conclusions are drawn in Section~\ref{sec:concl}.

\section{Two-fluid turbulent plasma model and its numerical implementation in GBS}
\label{sec:two-fluid}
\begin{figure}
    \centering
    \begin{subfigure}{0.48\textwidth}
        \includegraphics[width=\linewidth]{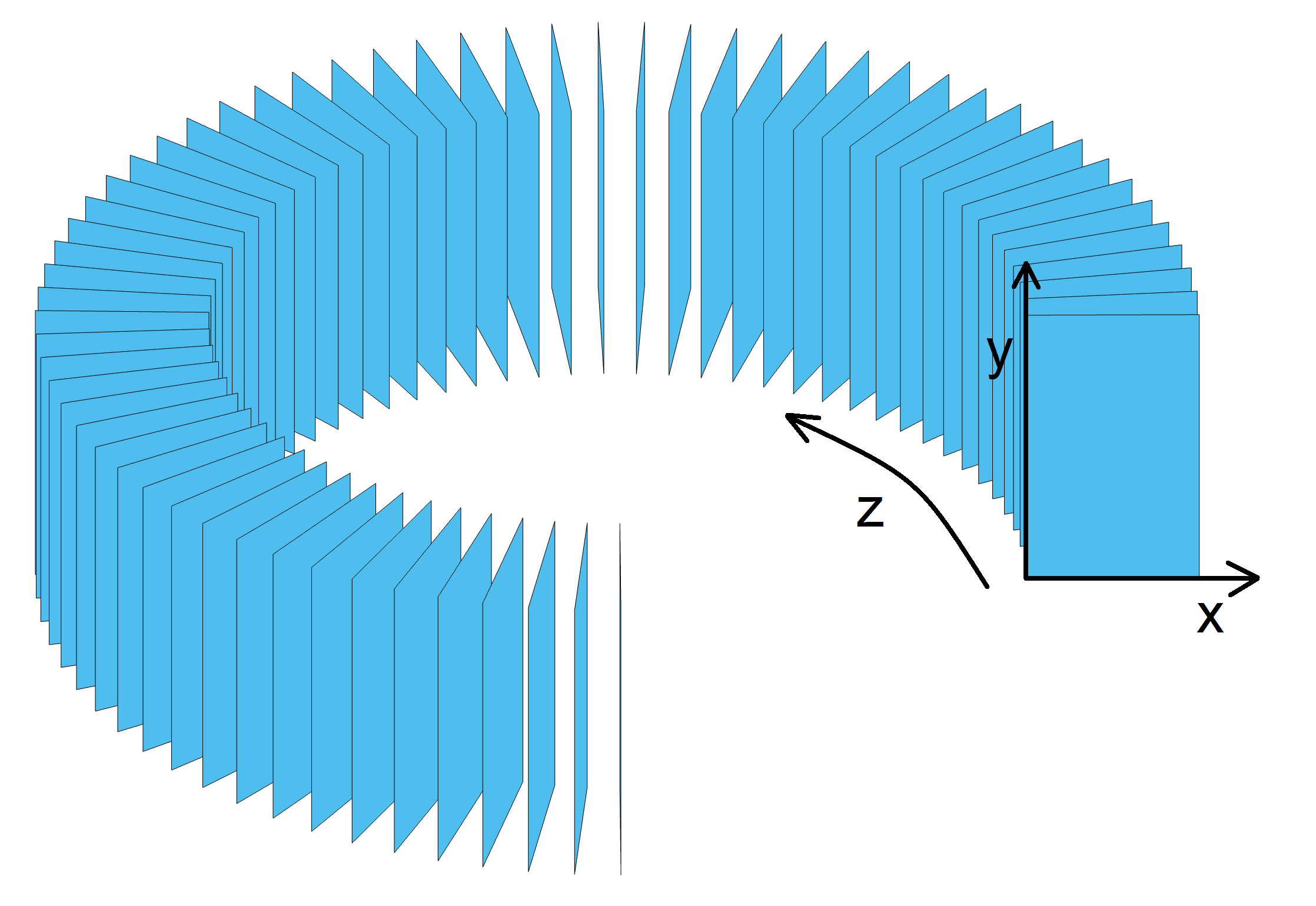}
          \caption{}
    \label{fig:domain}
    \end{subfigure}
    \begin{subfigure}{0.48\textwidth}
       \includegraphics[width=0.8\linewidth]{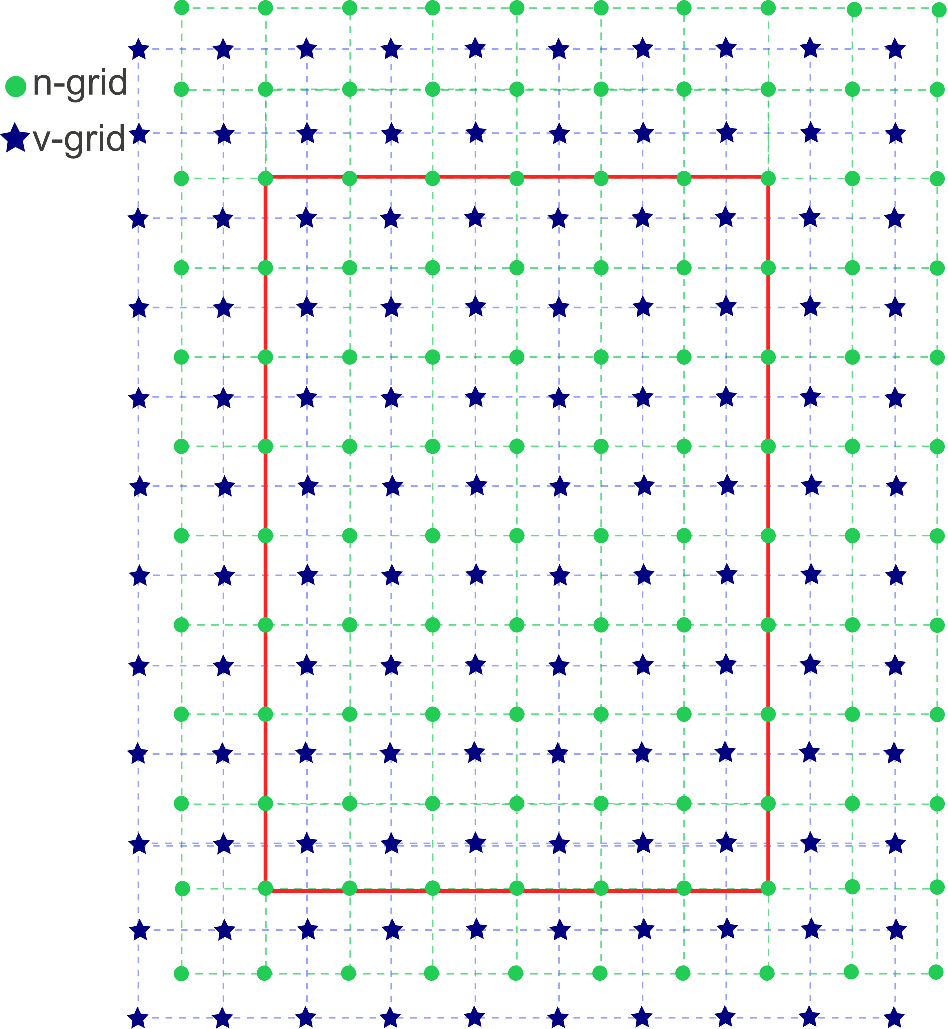}
    \caption{}
    \label{fig:grids}
    \end{subfigure}
    \caption{Sketch of the 3D computational domain including the system of coordinates $(x,y,z)$ in Fig.~\ref{fig:domain} and the discretization grid $n$-grid and $v$-grid in a 2D space in Fig.~\ref{fig:grids}. }
\end{figure}

The plasma boundary dynamics are modeled using the drift-reduced Braginskii equations \cite{braginskii1965transport}, derived, for example, in Zeiler et al. \cite{zeiler1997nonlinear}, as implemented in the GBS code \cite{giacomin2022gbs,ricci2012simulation,halpern2016gbs}. GBS is a three-dimensional, flux-driven, global, two-fluid turbulence code developed for the self-consistent simulation of plasma turbulence and neutral dynamics, described by using a kinetic model, in the tokamak boundary region \cite{wersal2015first}. In the present work, we restrict ourselves to the electrostatic limit and neglect interactions with neutrals. The governing dimensionless equations for plasma density $n$, scalar vorticity $\Omega$, electron and ion parallel velocities, $\vpare$ and $\vpari$, electron and ion temperatures, $T_e$ and $T_i$, read as follows:
\begin{align}
\label{eqn:density}
 \frac{\partial n}{\partial t} =& -\frac{\rho_*^{-1}}{B}[\phi,n]+\frac{2}{B}\Bigl[C(p_e)-nC(\phi)\Bigr] 
-\nabla_{\parallel}(n V_{\parallel e}) + D_n\nabla_{\perp}^2 n +s_n\, ,\\
\label{eqn:vorticity}
\frac{\partial \Omega}{\partial t} =& -\frac{\rho_*^{-1}}{B}\nabla \cdot [\phi,\boldsymbol{\omega}] - \nabla \cdot \bigl( V_{\parallel i}\nabla_\parallel \boldsymbol{\omega}\bigr) + B^2\nabla_{\parallel}j_{\parallel} + 2B C(p_e + \tau p_i) + \frac{B}{3}C(G_i) + D_{\Omega}\nabla_\perp^2 \Omega\, ,\\
\label{eqn:electron_velocity}
\frac{\partial V_{\parallel e}}{\partial t} =& -\frac{\rho_*^{-1}}{B}[\phi,V_{\parallel e}] - V_{\parallel e}\nabla_\parallel V_{\parallel e} + \frac{m_i}{m_e}\Bigl(\nu j_\parallel+\nabla_\parallel\phi-\frac{1}{n}\nabla_\parallel p_e-0.71\nabla_\parallel T_e \Bigr)-\frac{2}{3n}\nabla_\parallel G_e+ D_{V_{\parallel e}}\nabla_\perp^2 V_{\parallel e}\,, \\
\label{eqn:ion_velocity}
\frac{\partial V_{\parallel i}}{\partial t} =& -\frac{\rho_*^{-1}}{B}[\phi,V_{\parallel i}] - V_{\parallel i}\nabla_\parallel V_{\parallel i} - \frac{1}{ n}\nabla_\parallel(p_e+ \tau p_i) -\frac{2}{3n}\nabla_\parallel G_i+ D_{V_{\parallel i}}\nabla_\perp^2 V_{\parallel i}\, ,\\
\label{eqn:electron_temperature}
\frac{\partial T_e}{\partial t} =& -\frac{\rho_*^{-1}}{B}[\phi,T_e] - V_{\parallel e}\nabla_\parallel T_e 
+ \frac{2}{3}T_e\Bigl[0.71\frac{\nabla_\parallel j_\parallel}{n} - \nabla_\parallel V_{\parallel e}\Bigr] + \frac{4}{3}\frac{T_e}{B}\Bigl[\frac{7}{2}C(T_e)+\frac{T_e}{n}C(n)-C(\phi)\Bigr] 
 \nonumber \\
&+\frac{2}{3n}\nabla_\parallel (\chi_{\parallel e}\nabla_\parallel T_e) + D_{T_e}\nabla_\perp^2 T_e  + s_{T_e}- 2.61 \nu n(T_e- \tau T_i)  \,,\\
\label{eqn:ion_temperature}
\frac{\partial T_i}{\partial t} =& -\frac{\rho_*^{-1}}{B}[\phi,T_i] - V_{\parallel i}\nabla_\parallel T_i 
+ \frac{4}{3}\frac{T_i}{B}\Bigl[C(T_e)+\frac{T_e}{n}C(n)-C(\phi)\Bigr] - \frac{10}{3}\tau\frac{T_i}{B}C(T_i) \nonumber \\ 
&+ \frac{2}{3}T_i\Bigl[(V_{\parallel i}-V_{\parallel e})\frac{\nabla_\parallel n}{n} -\nabla_\parallel V_{\parallel e}\Bigr] 
 + \frac{2}{3n}\nabla_\parallel (\chi_{\parallel i}\nabla_\parallel T_i) + D_{T_i}\nabla_\perp^2 T_i + s_{T_i} + 2.61 \nu n (T_e- \tau T_i) \,,
\end{align}
and are coupled to the Poisson equation, for the electrostatic potential $\phi $
\begin{equation}
\nabla \cdot \bigl( n \nabla_\perp \phi\bigr) =\ \Omega-\tau \nabla_\perp^2 p_i.
\label{eqn:poisson}
\end{equation}
In Eqs.~\eqref{eqn:density}--\eqref{eqn:poisson}, the electron and ion pressures are denoted as $p_e = n T_e$ and $p_i = n T_i$, the dimensionless current is $j_{\parallel} = n(\vpari -\vpare)$, and $\Omega = \nabla\cdot\boldsymbol{\omega} = \nabla \cdot (n \nabla_\perp\phi + \nabla_\perp \tau p_i)$. The dimensionless parameter $\rho_*^{-1}$ is the ratio between the tokamak major radius and the ion sound Larmor radius; $\tau$ is the ion-to-electron reference temperature ratio; $\nu=\nu_0 T_e^{-3/2}$ is the normalized Spitzer resistivity; and $\chi_{\parallel e}$ and $\chi_{\parallel i}$ are the normalized electron and ion parallel thermal conductivities. The perpendicular diffusion terms $D_f\nlaplperp f$ are added for numerical stability.
The gyroviscous terms $G_e$ and $G_i$ are approximated by constant parallel diffusion terms for the velocities, $G_e=-2\eta_{0e}\ngradpar \vpare$ and $G_i=-2 \eta_{0i} \ngradpar \vpari$, where $\eta_{0e}$ and $\eta_{0i}$ are the electron and ion viscosities.
The spatial operators appearing in Eqs.~\eqref{eqn:density}–\eqref{eqn:poisson} are defined as follows. The Poisson bracket $[\phi,f]$ represents the $\mathbf{E}\times\mathbf{B}$ convective terms; the curvature operator $C(f)$ accounts for magnetic field line curvature effects; the parallel gradient $\ngradpar$ and parallel Laplacian $\nlaplpar$ describe transport and diffusion along the equilibrium magnetic field direction; and the perpendicular Laplacian $\nlaplperp$ describes diffusion in the plane perpendicular to the magnetic field. These operators are:
\begin{subequations}
\begin{align}
    [\phi,f]=&\mathbf{b}\ \cdot\ \bigl(\nabla \phi \times \nabla f\bigr)\,,\\
 C(f)=&\frac{B}{2}\Bigl(\nabla \times \frac{\mathbf{b}}{B}\Bigr)\cdot \nabla f\,,\\
  \nabla_\parallel f=&\mathbf{b}\cdot\nabla f \,,\\
  \nabla_\parallel^2 f=&\mathbf{b}\cdot\nabla(\mathbf{b}\cdot\nabla f) \,,\\
     \nabla_\perp^2 f=&\nabla\cdot\bigl[(\mathbf{b}\times\nabla f)\times\mathbf{b}\bigr]\,,
\end{align}
\label{eqn:operators}%
\end{subequations}
where $f$ is an arbitrary scalar function and $\mathbf{b}=\mathbf{B}/{B}$ the unit vector aligned with the magnetic field $\mathbf{B}$. The magnetic field is expressed in terms of the axisymmetric poloidal magnetic flux $\Psi$ as
\begin{equation}
 \mathbf{B}=R B_\varphi \nabla \varphi + \nabla \varphi \times \nabla \Psi,
    \label{eqn:mag_field}
\end{equation}
where $B_\varphi$ is the toroidal field strength and $\varphi$ is the toroidal angle. 
The operators in Eqs.~\eqref{eqn:operators} are expanded in cylindrical
coordinates $(Z,R, \varphi)$ and further simplified under the
large-aspect-ratio assumption, i.e., that the tokamak minor radius is much
smaller than the major radius, and that the magnetic field is predominantly
toroidal. Under these assumptions, the curvature of the torus
can be neglected at leading order, so the coordinates $(Z,R, \varphi)$ are treated as Cartesian coordinates $(y,x,z)$. The plasma domain is then approximated as a Cartesian volume $(y,x,z) \in \left[Y_{\min},Y_{\max}\right]\times
 \left[R_{\min},R_{\max}\right] \times \left[0,2\pi\right]$, with a periodic boundary condition in the $z$ direction. A schematic representation of the domain and the associated coordinate system is shown in Fig.~\ref{fig:domain}. The operators then reduce to:
\begin{subequations}
\begin{align}
    [\phi,f]=& b_{z} (\partial_y \phi \partial_x f- \partial_x \phi \partial_y f),\\
     C(f)=& b_z \partial_y f,\\
     \ngradpar f=& \partial_y \Psi\partial_x f -\partial_x \Psi \partial_y f+ b_z \partial_z f= \mathbf{\Tilde{b}}\cdot \nabla f,\\
     \nlaplpar f=& \nabla \cdot (\mathbf{\Tilde{b}\Tilde{b}^T}\nabla f),\\
     \nlaplperp f =& \partial_{xx}^2 f + \partial_{yy}^2 f,
\end{align}
\end{subequations}
where $\mathbf{\Tilde{b}}= \left[- \partial_x \Psi, \partial_y \Psi, b_z \right]$ then defines the direction of the magnetic field. 

Following the approach adopted by several other fluid turbulent plasma codes \cite{fevrier2024splend1d,Zholobenko_2021}, the parallel heat conductivities are treated using a flux-limiting formulation that interpolates between the Spitzer-Härm form and the free-streaming heat flux. The resulting effective electron and ion heat conductivities are given by
\begin{subequations}
\begin{align}
    \mathrm{\chi}_{\parallel e}= &\Tilde{\chi}^e_{\parallel 0} T_e^{5/2} \bigg(1+ \Tilde{\chi}^e_{\parallel 0}\sqrt{\frac{m_e}{m_i}}\frac{T_e^2}{\alpha_e n q }\bigg)^{-1},\\
    \mathrm{\chi}_{\parallel i}= &\Tilde{\chi}^i_{\parallel 0} T_i^{5/2} \bigg(1+ \Tilde{\chi}^i_{\parallel 0} \sqrt{\tau}\frac{T_i^2}{\alpha_i n q }\bigg)^{-1},
\end{align}
\label{eqn:flux_temp}%
\end{subequations}
where the safety factor $q$, which typically ranges between $2.5$ and $4$ for tokamaks, is set to $2.5$ in our simulations. The coefficients $\Tilde{\chi}^e_{\parallel 0}$ and $\Tilde{\chi}^i_{\parallel 0}$ are defined in App.~\ref{sec:app_therm}. Decreasing the parameters $\alpha_e$ and $\alpha_i$ reduces the effective thermal conductivities. In the present simulations, we adopt $\alpha_e=\alpha_i= 0.8$. Accounting for the temperature dependence of the thermal conductivity coefficients significantly increases the stiffness of the parallel conduction terms. The implementation of an IMEX scheme, which treats these stiff terms implicitly, now makes it possible to incorporate this dependence in a numerically tractable way in GBS.
A simplified version of the generalized Bohm-Chodura boundary conditions \cite{stangeby2000plasma}, derived following \cite{giacomin2022gbs,loizu2012boundary,mosetto2015finite}, is imposed at the magnetic pre-sheath entrance. More precisely, the boundary conditions imposed have the following form:
\begin{subequations}
\begin{align}
\partial_s T_e &=\partial_s T_i =0,\\
\vpari&=\vpare= \pm c_s \sqrt{1+ \tau\frac{T_i}{T_e}}, \\ 
\partial_s n&=\mp \frac{n}{c_s \sqrt{1+ \tau\frac{T_i}{T_e}}}\partial_s \vpari,\\
\Omega &= \mp {n c_s}\sqrt{1+\tau\frac{T_i}{T_e}}\partial_{ss}^2 \vpari,\\ 
\phi&=(\Lambda_0-\log \sqrt{1+\tau\frac{T_i}{T_e}})T_e, 
\end{align}
\label{eqn:bc}
\end{subequations}
where $\Lambda_0 =\log \sqrt{m_i/(2\pi m_e)}$ and $s$ denotes the coordinate normal to the wall, with $s=y$ for the top and bottom boundaries and $s=x $ for the inner and outer wall. To prevent discontinuities in the parallel velocities at locations where the magnetic field is tangent to the wall, a smoothing function is applied to transition the velocities from $+c_s$ to $-c_s$, where $c_s=\sqrt{T_e}$ is the normalized ion sound speed. This procedure ensures that $\vpari$ and $\vpare$ vary smoothly and remain free of sharp discontinuities.


To solve numerically the system of Eqs.~\eqref{eqn:density}--\eqref{eqn:poisson} spatial differential operators are discretized using centered finite difference schemes. To prevent the emergence of checkerboard patterns when using centered finite difference schemes \cite{BASSANINI2026114539}, a staggered grid arrangement is adopted. Specifically, two grids are employed: the $n$-grid and the $v$-grid, which are offset from each other by half a cell in all spatial directions as shown in 2D in Fig.~\ref{fig:grids}. The scalar quantities $n$, $\phi$, $\Omega$, $T_e$, and $T_i$ are defined on the $n$-grid, while the parallel velocities $\vpare$ and $\vpari$ are defined on the $v$-grid.

The parallel Laplacian operator $\nlaplpar$ is discretized using the second-order symmetric scheme proposed in \cite{gunter2005modelling}, which relies on a $3 \times 3 \times 3$ stencil in three dimensions. For the parallel gradient operator $\ngradpar$, which maps fields between the two staggered grids, we employ the mimetic finite difference approach introduced in \cite{BASSANINI2026114539}. The Poisson bracket is discretized using the Arakawa scheme \cite{peterson2013positivity}. Moreover, to preserve the positivity of the temperatures and the density, we evolve their logarithms, defining $t_e=\ln{T_e}$, $t_i=\ln{T_i}$, and $\theta=\ln{n}$; the corresponding evolution equations are obtained by dividing Eqs.~\eqref{eqn:density}, \eqref{eqn:electron_temperature}, and \eqref{eqn:ion_temperature} by $n$, $T_e$, and $T_i$, respectively.

GBS is parallelized using the Message Passing Interface (MPI) with a Cartesian communicator, where each process is assigned a subdomain with a cubic structure consistent with the spatial discretization. To support the evaluation of fourth-order finite difference stencils, each subdomain includes two ghost cells in every spatial direction. The only exception is the $\nlaplpar$ operator, which is discretized using a second-order scheme, to avoid the three ghost cells per direction that a fourth-order operator would require, as in \cite{gunter2007finite}.

\section{Implementation of an IMEX scheme in GBS}
\label{sec:imex}
The time integration scheme employed in this work belongs to the family of Implicit-Explicit Runge-Kutta (IMEX-RK) methods, specifically the class of additive Runge-Kutta schemes first introduced by Ascher et al.\ \cite{ascher1997implicit} in the context of the numerical solution of convection-diffusion equations. A representative example of the applicability of such methods is found in non-hydrostatic atmospheric models \cite{gardner2018implicit}, where the dynamics naturally unfolds across multiple timescales: fast vertical waves that require implicit treatment, fast horizontal waves handled through explicit substepping relative to the remaining dynamics, and slow large-scale processes advanced explicitly with a larger time step.

The construction of an effective IMEX-RK scheme requires a careful identification and separation of the terms present in the right-hand side (RHS) of the governing equations according to the physical processes they represent and the characteristic timescales that they act upon. 
When applied to a differential-algebraic equation (DAE) system of index 1 \cite{wanner1996solving, roche1989ernst} of the form
\begin{equation}
\begin{aligned}
\pdv{\textbf{Q}}{t} &= \textbf{F}(\textbf{Q},\zeta) \\
0 &= g(\textbf{Q},\zeta)
\end{aligned}
\label{eqn:dae_1}
\end{equation}
where $\textbf{Q}$ is the state vector of the system and $\zeta$ is the algebraic variable determined by the constraint $g$, the central idea of the IMEX-RK methods is to split the RHS operator $\textbf{F}(\textbf{Q},\zeta)$ into two contributions, that is  
\begin{equation}
    \textbf{F}(\textbf{Q},\zeta)=\textbf{F}_F(\textbf{Q}, \zeta )+\textbf{F}_S(\textbf{Q},\zeta).
    \label{eqn:fast_vs_slow}
\end{equation}
The stiff part, $\textbf{F}_F(\textbf{Q},\zeta)$, associated with the fast phenomena, is treated implicitly; the non-stiff part, $\textbf{F}_S(\textbf{Q},\zeta)$, associated with the slow or moderately varying dynamics, is treated explicitly \cite{kennedy2003additive}. This splitting strategy ensures that the implicit solve is restricted to the stiff subsystem only, which is typically simpler in structure and therefore more tractable for efficient inversion. Meanwhile, the explicit treatment of the non-stiff terms recovers the computational simplicity and parallelization efficiency of explicit methods. The resulting CFL condition is governed solely by the non-stiff terms and is therefore far less restrictive than that of a fully explicit scheme, enabling the use of significantly larger time steps without sacrificing stability. The overall efficiency of the IMEX-RK approach thus stems from the balance between stability requirements and computational cost, making it particularly well-suited for the class of problems considered in the present work.

If the IMEX-RK methods applied to a DAE system is globally stiffly accurate (GSA), which means that the last internal stage coincides with the solution at the next time step \cite{boscarino2013implicit}, it is also $L$-stable \cite{boscarino2024implicit}. The $L$-stability property is particularly desirable for stiff problems, where rapidly decaying transient components may be present in the solution, as it guarantees that such transients are effectively damped within a single time step. This enables the use of time steps far larger than those that would otherwise be required to accurately resolve the transient phase \cite{boscarino2024implicit}. Moreover, for computational efficiency, the implicit part of the scheme is usually chosen to be a diagonally implicit Runge–Kutta (DIRK) method.

The implementation of the GSA $s$-stage IMEX scheme (where $s$ denotes the number of internal stages) applied to the DAE system~\eqref{eqn:dae_1} reads:
\begin{align*}
&\begin{cases}
\displaystyle
    \textbf{Q}^{(i)} =\textbf{Q}^n+\Delta t \sum_{j=1}^{i-1}a_{ij}\textbf{F}_S\left(\textbf{Q}^{(j)},\zeta^{(j)}\right)+\Delta t \sum_{j=1}^{i}\Bar{a}_{ij}\textbf{F}_F\left(\textbf{Q}^{(j)},\zeta^{(j)}\right),\\
0=g(\textbf{Q}^{(i)},\zeta^{(i)}),
\end{cases} \text{ for }i=1,...,s
\end{align*}
where the GSA property directly yields the solution at the next time step, $\textbf{Q}^{n+1}= \textbf{Q}^{(s)}$ and $ \zeta^{n+1}=\zeta^{(s)}$ without requiring an additional update step. The coefficients $a_{ij}$ and $\Bar{a}_{ij}$ form the explicit matrix $A$ and the implicit matrix $\Bar{A}$, respectively, which together constitute the \textit{Butcher tableaux} of the IMEX-RK scheme.

At each internal stage, the IMEX structure of the scheme requires the solution of a system of the following form:
\begin{equation}
\begin{cases}
      \textbf{Q}^{(i)}-\Delta t \Bar{a}_{ii}\textbf{F}_F\left(\textbf{Q}^{(i)},\zeta^{(i)}\right)=\textbf{K}^{(i)},\\
      0=g(\textbf{Q}^{(i)},\zeta^{(i)}),
\end{cases} 
\label{eqn:system_to_besolved}
\end{equation}
where the right-hand side $\textbf{K}^{(i)}$ is defined as a linear combination of the slow and fast terms evaluated at all previous internal stages:
\begin{equation}
    \textbf{K}^{(i)}\coloneqq \textbf{Q}^n+\Delta t \sum_{j=1}^{i-1}a_{ij}\textbf{F}_S\left(\textbf{Q}^{(j)},\zeta^{(j)}\right)+\Delta t \sum_{j=1}^{i-1}\Bar{a}_{ij}\textbf{F}_F\left(\textbf{Q}^{(j)},\zeta^{(j)}\right).
    \label{eqn:rhs_systemimex}
\end{equation}

Among the large family of $s$-stage IMEX-RK schemes, we employ the third-order, globally stiffly accurate (GSA) BPR(3,4,3) method \cite{boscarino2013implicit}, where the set (3,4,3) denotes, respectively, the number of explicit internal stages, the number of implicit internal stages, and the overall order of accuracy of the scheme. Furthermore, BPR(3,4,3) adopts a singly diagonally implicit Runge–Kutta (SDIRK) structure for its implicit part, which is particularly advantageous when using iterative solvers, as the same operator structure is reused across all stages, reducing both setup cost and memory requirements. The matrices $A$ and $\Bar{A}$ for the BPR(3,4,3) scheme are reported in App.~\ref{sec:app_butcher}.

The system reported in Eqs.~\eqref{eqn:density}-\eqref{eqn:poisson} can be written in compact form as Eq.~\eqref{eqn:dae_1}
where $\textbf{Q}=\left[ n,\Omega, \vpare, \vpari, T_e, T_i\right]^T$ collects the time-dependent unknowns, the algebraic variable is the electrostatic potential $\zeta=\phi$ and $g$ denotes the Poisson equation in \eqref{eqn:poisson}. The drift-reduced Braginskii equations exhibit several phenomena occurring on fast timescales, which motivate the splitting of the right-hand side into stiff and non-stiff contributions. 
A first source of stiffness is the electrostatic Shear Alfvén wave (SAW)~\cite{Jolliet:207684, BASSANINI2026114539, STEGMEIR2023108801}. It can be isolated by assuming that the density and temperatures vary slowly, and may therefore be treated as constant values $n_0, T_{e0}, T_{i0}$, while neglecting the ion parallel velocity. These assumptions yield the following reduced subsystem governing the electron parallel velocity and the vorticity:
\begin{equation}
    \begin{aligned}
    \pdv{\vpare}{t}=&\frac{m_i}{m_e} \ngradpar \phi  + \frac{4\eta_{0e}}{3n_0}\nlaplpar \vpare,\\
    \pdv{\Omega}{t}=& -n_0\ngradpar \vpare,
    \end{aligned}
    \label{eqn:saws_imp}
\end{equation}
subject to the algebraic constraint $n_0\nlaplperp \phi=\Omega$ for the electrostatic potential.
The dispersion relation of the SAWs \cite{BASSANINI2026114539} reveals a wave frequency of the form $\omega= -i\gamma_D\pm \sqrt{\omega_0^2-\gamma_D^2}$ where 
\begin{align}
&\omega_0= \sqrt{\frac{m_i}{ m_e}\frac{k_\parallel^2}{k_\perp^2}},\label{eqn:estimation_frequency}\\
&\gamma _D=\frac{2\eta_{0e} }{3n_0}k_\parallel^2.
\end{align}
Since $\omega_0 \propto k_\parallel/k_\perp$, the fastest SAW oscillation, defining the CFL condition, corresponds to the largest parallel wavenumber and the smallest perpendicular wavenumber that the grid can support. The largest resolvable parallel wavenumber is set by the grid spacing in the $z$ direction, $k_\parallel \sim N_z$, where $N_z$ is the number of planes along $z$, while the smallest perpendicular wavenumber corresponds to the longest wavelength fitting in the domain, $k_\perp \sim 2\pi/L_y$, where $L_y$ is the maximum domain size (usually along the $y$ direction). Substituting these estimates yields the maximum SAW frequency, $\omega_0\sim \sqrt{m_i/ m_e}(N_z L_y)/(2\pi)$.
The parallel diffusion introduces a damping rate $\gamma_D$, which also reduces the oscillation frequency. The stability constraint imposed by the SAWs on an explicit time stepping scheme is given by $\Delta t \lesssim (2\pi)/(L_y N_z)\sqrt{m_e/m_i}$. In the electrostatic limit considered here, the SAW dynamics become very fast for small perpendicular wavenumbers and, consistent with the extensive analysis of Ohm's law in reduced fluid models reported by \citeauthor{dudson2021ohm}~\cite{dudson2021ohm}, the characteristic speed of this fast wave grows rapidly with system size, making its implicit treatment increasingly critical at larger scales. 
We note that, when the drift-reduced Braginskii equations incorporate electromagnetic fluctuations, SAWs still impose a restrictive CFL constraint on explicit time stepping; this constraint, however, no longer grows unboundedly with system size.


The second source of stiffness arises from parallel electron and ion heat conduction along the magnetic field lines:
\begin{equation}
    \pdv{T_{a}}{t}=\frac{2}{3n}\ngradpar(\chi_{\parallel a}\ngradpar T_{a}),
    \label{eqn:temp_imp}
\end{equation}
where $a\in\{e,i\}$ denotes electrons or ions \cite{dudson2012improved}. Additionally, the electron and ion viscosity terms are also found to introduce numerical stiffness:
\begin{equation}
    \pdv{V_{\parallel a}}{t}= \frac{4\eta_{0a}}{3n}\nlaplpar V_{\parallel a}.
    \label{eqn:v_imp}
\end{equation}

The fast, stiff terms identified in Eqs.~\eqref{eqn:saws_imp}, \eqref{eqn:temp_imp} and \eqref{eqn:v_imp} are collected in $\textbf{F}_F$ in Eq.~\eqref{eqn:fast_vs_slow}, while $\textbf{F}_S$ gathers the remaining slow, non-stiff terms. In contrast to the SAW subsystem, the stiff subsystems associated with parallel heat conduction and ion viscosity each involve a single unknown with no coupling to other variables. This significantly simplifies the construction of an efficient preconditioner. The only exception is the electron parallel velocity $\vpare$, whose viscosity term is coupled to the SAW system and must therefore be treated as part of that coupled subsystem rather than independently.

At each internal stage $i \in \{1,\dots, s\}$ of the IMEX scheme, we compute the right-hand sides $\text{K}_g^{(i)}$ of the linear systems associated with each variable $g \in \{T_e, T_i, \vpari, \vpare, \Omega\}$, according to Eq.~\eqref{eqn:rhs_systemimex}, and for $\phi$ according to the algebraic constraint in Eq.~\eqref{eqn:poisson}.
Since the boundary conditions of all variables depend on the values of other variables at the boundary except for the temperatures, a specific ordering on the solution of the linear systems should be imposed.

The systems of equations for the electron and ion temperatures are solved first, since their boundary conditions do not depend on any other variable. The linear system to be solved for the electron temperature at each internal stage reads:
\begin{equation}
\left[\begin{matrix} \textbf{I}-\Bar{a}_{ii}\Delta t \dfrac{2}{3n^{(i-1)}T_e^{(i-1)}}\ngradpar \big(\chi_{\parallel e }^{(i-1)}T_e^{(i-1)}\ngradpar\big)\end{matrix}\right]\left[\begin{matrix}t_e^{(i)}\end{matrix}\right]=\left[\begin{matrix}\text{K}_{\tempe}^{(i)}\end{matrix}\right],
    \label{eqn:tempe_imp}
\end{equation}
where the coefficients of the diffusion operator are linearized by evaluating them at the previous internal stage $(i-1)$ of the IMEX scheme, avoiding the need to solve a nonlinear system at each stage. An analogous system is obtained for the ion temperature:
\begin{equation}
\left[\begin{matrix} \textbf{I}-\Bar{a}_{ii}\Delta t  \dfrac{2}{3n^{(i-1)}T_i^{(i-1)}}\ngradpar \big(\chi_{\parallel i}^{(i-1)} T_i^{(i-1)} \ngradpar\big)\end{matrix}\right]\left[\begin{matrix}t_i^{(i)}\end{matrix}\right]=\left[\begin{matrix}\text{K}_{\tempi}^{(i)}\end{matrix}\right],
    \label{eqn:tempi_imp}
\end{equation}
where the same linearization strategy is applied.
Once Eqs.~\eqref{eqn:tempe_imp} and \eqref{eqn:tempi_imp} are solved, the ion parallel velocity can be advanced, since its boundary conditions depend only on the temperature values. The corresponding linear system reads:
\begin{equation}
\left[\begin{matrix} \textbf{I}-\Bar{a}_{ii}\Delta t \dfrac{4\eta_{0i}}{3n^{(i-1)}}\nlaplpar \end{matrix}\right]\left[\begin{matrix}\vpari^{(i)}\end{matrix}\right]=\left[\begin{matrix}\text{K}_{\vpari}^{(i)}\end{matrix}\right],
\label{eqn:vpari_imp}
\end{equation}
where the density is evaluated at the previous internal stage $(i-1)$.
The density is then advanced explicitly as:
\begin{equation}
\theta^{(i)} =\theta^n+\Delta t \sum_{j=1}^{i-1}a_{ij} F_{\theta,S} 
    \label{eqn:density_imp}
\end{equation}
since the density equation does not contain fast stiff terms and therefore does not require implicit treatment.
Having the value of the density at stage $(i)$, the coupling between the SAW dynamics, Eq.~\eqref{eqn:saws_imp}, and the electron parallel viscosity yields a linear system for the vorticity $\Omega^{(i)}$, the electron parallel velocity $\vpare^{(i)}$, and the electrostatic potential $\phi^{(i)}$:
\begin{equation}
\mathbb{A}\textbf{v}=\textbf{r}
\label{eqn:imex}
\end{equation}
with 
\begin{equation}
\mathbb{A}=\left[\begin{matrix} \textbf{I}& \Bar{a}_{ii}\Delta tn^{(i)}\ngradpar & \textbf{0} \\[5pt] \textbf{0} & \textbf{I} -\Bar{a}_{ii}\Delta t \frac{4\eta_{0e}}{3n^{(i)}} \nlaplpar & -\Bar{a}_{ii} \Delta t \frac{m_i}{m_e}\ngradpar \\[5pt] \textbf{I} &\textbf{0}& -\nabla\cdot (n^{(i)} \ngradperp) \end{matrix}\right], \textbf{v}=\left[\begin{matrix}
        \Omega^{(i)}\\[5pt] \vpare^{(i)}\\[5pt] \phi^{(i)}
    \end{matrix}\right], \textbf{r}=\left[\begin{matrix}
        \text{K}_{\Omega}^{(i)}\\[7pt]
        \text{K}_{\vpare}^{(i)}\\[7pt]
        \text{K}_{\phi}^{(i)}
    \end{matrix}\right],
\end{equation}
where $\text{K}_g^{(i)}$ for $g \in \{\Omega, \vpare\}$  corresponds to the row of Eq.~\eqref{eqn:rhs_systemimex} associated with variable $g$, and $\text{K}_\phi^{(i)}$ is defined as 
\begin{equation}
    \text{K}_\phi^{(i)}= \tau \nlaplperp p_i^{(i)}.
    \label{eqn:phi_rhs}
\end{equation}
This system in Eqs.~\eqref{eqn:imex} is solved using the physics-based preconditioning strategy described in Sec.~\ref{sec:pb_prec}. The complete solution procedure used at each internal stage of the adopted IMEX-RK method is summarized in Alg.~\ref{alg:1}.

\begin{algorithm}[t]
\linespread{1.3}\selectfont
\caption{Solution of one time step with an $s$-stage IMEX-RK}
\label{alg:1}

\begin{algorithmic}[1]

\For{$i = 1, \dots, s$}

    \State Compute RHS as defined in Eq.~\eqref{eqn:rhs_systemimex} and in Eq.~\eqref{eqn:phi_rhs}:
    $\text{K}_{\tempe}^{(i)}, \text{K}_{\tempi}^{(i)}, \text{K}_{\vpari}^{(i)}, 
    \text{K}_{\Omega}^{(i)}, \text{K}_{\vpare}^{(i)}, \text{K}_{\phi}^{(i)}$

    \State Solve Eq.~\eqref{eqn:tempe_imp} and \eqref{eqn:tempi_imp} with GMRES using 3D AMG
    \State Solve Eq.~\eqref{eqn:vpari_imp} with GMRES using 3D AMG
    \State Explicit advancement of the density as in Eq.~\eqref{eqn:density_imp}
    \State Solve Eq.~\eqref{eqn:imex} with FGMRES using a physics-based preconditioner 
    \label{line:physics-b}

\EndFor
\end{algorithmic}
  \end{algorithm}

For the linear systems in Eqs.~\eqref{eqn:tempe_imp}--\eqref{eqn:vpari_imp}, \eqref{eqn:imex}, the initial guess is set to the solution obtained at the previous internal stage, and the algebraic preconditioners are computed and reused across subsequent stages until the number of iterations of the corresponding solver exceeds a prescribed threshold. This reuse strategy avoids the overhead of recomputing the preconditioner at every stage, while the threshold ensures that the preconditioner is updated whenever convergence degrades. 

Focusing on the implementation, we note that the mutual independence of the two temperature equations permits their solution in parallel. The available processes are split into two groups solving the electron and ion temperature systems simultaneously. However, this improves performance only when the efficiency gained by solving the systems in parallel on fewer ranks outweighs the performance loss from halving the rank count.
It also requires the two solves to be well load-balanced. In contrast, all other variables exhibit a cascade dependency: the solution of each subsystem depends on the boundary conditions provided by the previously solved one, preventing their concurrent execution. Indeed, as can be seen from Eqs.~\eqref{eqn:bc}, the boundary conditions of the temperatures are independent of all other variables; those of the ion parallel velocity depend on the temperatures; and those of the remaining variables in Eqs.~\eqref{eqn:imex} depend on the other variables solved beforehand.

\section{Physics-based preconditioner}
\label{sec:pb_prec}
The three linear systems for the temperatures and the ion parallel velocity in Eqs.~\eqref{eqn:tempe_imp}--\eqref{eqn:vpari_imp} share a diffusive structure, and are each solved using GMRES preconditioned by a three-dimensional algebraic multigrid (AMG) method. AMG preconditioners are widely employed for the iterative solution of large-scale sparse linear systems, particularly those arising from elliptic PDEs, owing to their optimal complexity, whereby the computational cost scales linearly with the problem size, and their robust convergence properties \cite{d2021amg}.
While the temperature and ion viscosity subsystems each involve a single variable avoiding coupling with other variables, the system of Eqs.~\eqref{eqn:imex} couples multiple variables through both differential operators and an algebraic constraint, making it the most challenging component to solve efficiently. We thus focus on the system of Eqs.~\eqref{eqn:imex} and construct a physics-based preconditioner it.

The system matrix $\mathbb{A}$ in Eq.~\eqref{eqn:imex} can be decomposed as the sum of two terms, $\mathbb{A}=\mathbb{B}+\mathbb{C}$, where the matrix 
\begin{equation*}
\mathbb{B}=\nolinebreak\left[\begin{matrix} 
\textbf{I}& \Bar{a}_{ii}\Delta tn^{(i)}\ngradpar & \textbf{0} \\[5pt] 
\textbf{0} & \textbf{I} & -\Bar{a}_{ii} \Delta t \frac{m_i}{m_e}\ngradpar \\[5pt] 
\textbf{I} &\textbf{0}& -\nabla\cdot (n^{(i)} \ngradperp) 
\end{matrix}\right]
\end{equation*}
encodes the SAW coupling between the three unknowns, and the matrix 
\begin{equation*}
    \mathbb{C}=\left[\begin{matrix}  \textbf{0} &\textbf{0} &\textbf{0} \\[5pt]\textbf{0} &   -\Bar{a}_{ii}\Delta t \frac{4\eta_{0e}}{3n^{(i)}} \nlaplpar &\textbf{0} \\[5pt] \textbf{0} &\textbf{0}&\textbf{0}\end{matrix}\right]
\end{equation*} accounts for the electron parallel viscosity.

This decomposition motivates the construction of a physics-based preconditioner, inspired by the parabolization strategy of \citeauthor{chacon2002implicit} in the MHD context~\cite{chacon2002implicit}, which approximates $\mathbb{A}$ by exploiting the structure of $\mathbb{B}$ and $\mathbb{C}$. Since this preconditioner involves inner iterative solves whose action varies between applications, a flexible variant of GMRES is required; accordingly, the system~\eqref{eqn:imex} is solved using FGMRES~\cite{saad1993flexible}.
The SAW coupling, captured by the matrix $\mathbb{B}$, and the parallel diffusion along the magnetic field lines, encoded in the operator $\mathbb{C}$, are treated separately in the preconditioner: the former through parabolization of the SAW block, following the key idea of \cite{chacon2002implicit}, the latter through an algebraic multigrid (AMG) preconditioner.

We begin by analyzing the structure of the SAW block $\mathbb{B}$ and derive its LDU block factorization $\mathbb{B}=\mathbb{L}\mathbb{D}\mathbb{U}$ where
\begin{equation*}
    \mathbb{L}=\left[\begin{matrix}
\textbf{I} &\textbf{0} & \textbf{0}\\[5pt]
\textbf{0} & \textbf{I} &\textbf{0}\\[5pt]
\textbf{I} & -\mathbf{T}_{\vpare} & -\textbf{I}  
\end{matrix}\right], \quad 
\mathbb{D}=\left[\begin{matrix} 
\textbf{I} &\textbf{0}&\textbf{0}\\[5pt]
\textbf{0} &\textbf{I}& \textbf{0}\\[5pt]
\textbf{0}&\textbf{0} & \mathbf{S}
\end{matrix}\right], \quad
\mathbb{U}=\left[\begin{matrix} 
\textbf{I}& \mathbf{T}_{\vpare}& \textbf{0} \\[5pt] 
\textbf{0} & \textbf{I} & \mathbf{T}_{\phi} \\[5pt] 
\textbf{0} &\textbf{0}& \textbf{I} 
\end{matrix}\right],
\end{equation*}
$\mathbb{L}$, $\mathbb{D}$, and $\mathbb{U}$ are lower block triangular, block diagonal, and upper block triangular matrices, respectively. 
The block operators appearing in this decomposition are defined as
\begin{align}
\mathbf{S}=&\nabla \cdot( n^{(i)} \ngradperp) 
+ \Bar{a}_{ii}^2\Delta t^2 \frac{m_i}{m_e}n^{(i)}\nlaplpar ,\\
\mathbf{T}_{\vpare}=&\Bar{a}_{ii}\Delta tn^{(i)}\ngradpar ,\\
\mathbf{T}_{\phi}=&-\Bar{a}_{ii} \Delta t \frac{m_i}{m_e}\ngradpar .
\end{align}
This decomposition is particularly valuable for preconditioner construction, as it isolates the dominant physical couplings into block operators that can be treated independently. In particular, the operator $\mathbf{S}$ corresponds to the Schur complement associated with the electrostatic potential block, and encodes the parabolic character introduced by the parabolization strategy.
To treat the electron viscosity contribution, we introduce the matrix $\widetilde{\mathbf{C}}$ associated with 
\begin{align}
\widetilde{\mathbf{C}}=\textbf{I} -\Bar{a}_{ii}\Delta t \frac{4\eta_{0e}}{3n^{(i)}} \nlaplpar ,
\end{align}
and define the preconditioner as
\begin{equation}
\mathbb{P}=\mathbb{B}\left[\begin{matrix}
\textbf{I} &\textbf{0} &\textbf{0} \\[5pt] 
\textbf{0} & \widetilde{\mathbf{C}} &\textbf{0} \\[5pt] 
\textbf{0} &\textbf{0} &\textbf{I}  
\end{matrix}\right] =
\mathbb{L}\mathbb{D}\mathbb{U}
\left[\begin{matrix}
\textbf{I} &\textbf{0} &\textbf{0} \\[5pt] 
\textbf{0} & \widetilde{\mathbf{C}} &\textbf{0} \\[5pt] 
\textbf{0} &\textbf{0} &\textbf{I}  
\end{matrix}\right]
\label{eqn:prec_inve}
\end{equation}
such that 
\begin{equation}
\mathbb{P}^{-1}=\left[\begin{matrix}
\textbf{I} &\textbf{0} &\textbf{0} \\[5pt] 
\textbf{0} & \widetilde{\mathbf{C}} &\textbf{0} \\[5pt] 
\textbf{0} &\textbf{0} &\textbf{I}  
\end{matrix}\right]^{-1}\mathbb{U}^{-1}\mathbb{D}^{-1}\mathbb{L}^{-1}.
\label{eqn:prec_inve2}
\end{equation}
The preconditioner $\mathbb{P}$, given in Eq.~\eqref{eqn:prec_inve}, provides an approximation of the matrix $\mathbb{A}$ and captures the dominant physical effects while maintaining a computationally efficient structure to invert.
Its application to the solution of system of Eqs.~\eqref{eqn:imex} reduces to the inversion of the Schur complement $\mathbf{S}$ and the diagonal block $\widetilde{\mathbf{C}}$ which, owing to their respective wave-like and diffusive character, are both well suited to iterative solution techniques accelerated by algebraic multigrid preconditioning.

The application of the physics-based preconditioner, that is the evaluation of $\mathbb{P}\textbf{x}^{out}=\textbf{x}^{in}$, at every iteration of the FGMRES method (line~\ref{line:physics-b} of Alg.~\ref{alg:1}), proceeds through the following sequence of operations:
\begin{enumerate}[label=(\roman*)]
    \item Given an input vector $\mathbf{x}^{in} = [\Omega^{in}, \vpare^{in}, \phi^{in}]^T$, we first solve (see Eq.~\eqref{eqn:prec_inve2}) the linear system:
    \begin{equation*}
        \mathbb{L} \mathbf{x}^{out,1} = \mathbf{x}^{in},
    \end{equation*}
   which in expanded form reduces to a single explicit evaluation of the electrostatic potential:

    \begin{equation*}
        \phi^{out,1} = \Omega^{in} - \phi^{in} - \mathbf{T}_{\vpare}\vpare^{in}.
    \end{equation*}

    \item The block matrix $\mathbb{D}$ is then applied to capture the shear
Alfvén wave dynamics, which requires solving
    \begin{equation}
        \mathbf{S} \phi^{out,2} = \phi^{out,1}.
        \label{eqn:wave}
    \end{equation}
    This system is solved using FGMRES preconditioned by an algebraic multigrid (AMG) method.

    \item Next, the upper block triangular matrix $\mathbb{U}$ is applied, leading to the sequential updates
    \begin{align*}
        \vpare^{out,3} &= \vpare^{in} - \mathbf{T}_{\phi}\phi^{out,2}, \\
        \Omega^{out,3} &= \Omega^{in} - \mathbf{T}_{\vpare}\vpare^{out,3}.
    \end{align*}

    \item Finally,  the contribution of electron parallel viscosity is incorporated by solving the diffusion problem
    \begin{equation}
        \widetilde{\mathbf{C}} \vpare^{out,4} = \vpare^{out,3},
        \label{eqn:diff}
    \end{equation}
     again using FGMRES preconditioned by a three-dimensional AMG method. The final output is $(\textbf{x}^{out})^T=\left[\Omega^{out,3},\vpare^{out,4},\phi^{out,2}\right].$
\end{enumerate}
Each application of the preconditioner for system~\eqref{eqn:imex} within a single iteration of the outer FGMRES solver therefore involves two inner FGMRES solvers: one associated with the parabolic SAW operator $\mathbf{S}$ (Eq.~\eqref{eqn:wave}) and one associated with the parallel diffusion operator $\widetilde{\mathbf{C}}$ (Eq.~\eqref{eqn:diff}).

Since the inner solvers, those solving Eqs.~\eqref{eqn:wave} and \eqref{eqn:diff}, operate on an approximation of the outer problem (Eqs.~\eqref{eqn:imex}), high accuracy is not required, and one or two AMG V-cycles are expected to suffice. However, since we reuse the AMG preconditioners across multiple FGMRES iterations of the inner solvers, their convergence is controlled by a relative tolerance of $10^{-1}$ (with respect to the right-hand side norm) rather than a fixed number of V-cycles. This tolerance-based criterion provides a natural and automatic indicator for deciding when the AMG preconditioners are updated, precisely, when the number of inner iterations required to reach the prescribed tolerance exceeds a given threshold, the preconditioner is recomputed. The AMG preconditioner employed in both inner solves is \textit{BoomerAMG} \cite{yang2002boomeramg} from the HYPRE library \cite{falgout2002hypre}, with the \textit{Falgout} coarsening strategy.

\section{Numerical results}
\label{sec:num}
In this section, we verify the implementation of the IMEX scheme in GBS (Sec.~\ref{sec:verification}) and evaluate its scalability via strong and weak scaling tests (Sec.~\ref{sec:parall}). We also compare the performance of the IMEX scheme with physics-based preconditioning against that of an adaptive explicit time integration scheme, measured in terms of wall clock time (Sec.~\ref{sec:efficiency}). The numerical tests are carried out on the CPU partition of LUMI (LUMI-C) \cite{lumi}. Each LUMI-C compute node is equipped with two AMD EPYC 7763 CPUs with 64 cores each running at 2.45 GHz for a total of 128 cores per node.
\begin{figure}[t]
    \centering
    \begin{subfigure}{0.47\textwidth}
        \centering
        \includegraphics[width=1.1\linewidth]{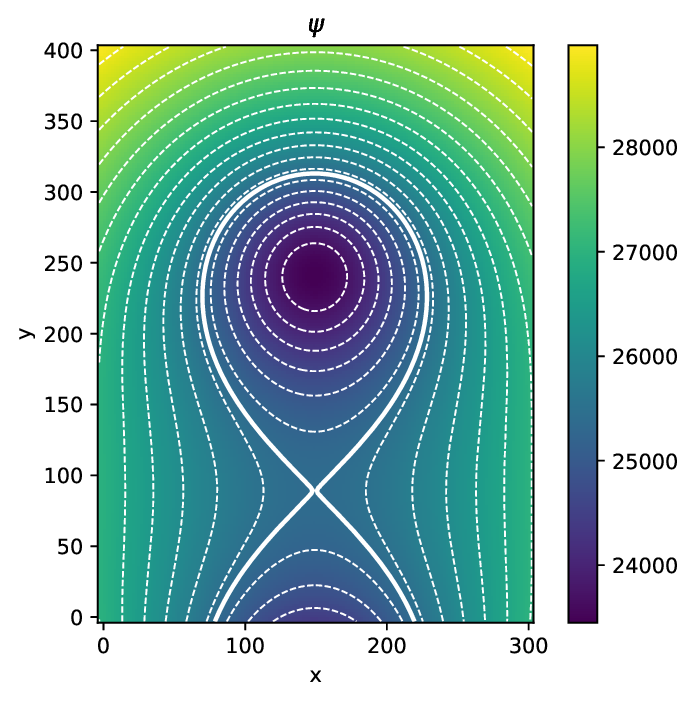}
        \caption{}
        \label{fig:equil_ana}
    \end{subfigure}
    \hfill
    \begin{subfigure}{0.47\textwidth}
        \hspace{8mm}
        \includegraphics[width=0.835\linewidth]{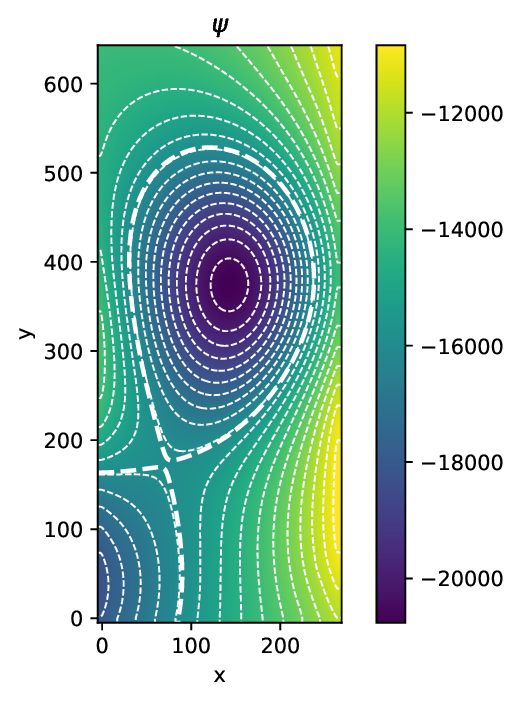}
        \caption{}
        \label{fig:equil_sperim}
    \end{subfigure}
    \caption{Contour plots of the poloidal magnetic flux $\Psi$ are presented for the analytical expression given by Eq.~\eqref{eqn:psi} in Fig.~\ref{fig:equil_ana}, where the separatrix is drawn as a solid white line, and for the experimental equilibrium in Fig.~\ref{fig:equil_sperim}, where the separatrix is shown as a wide dashed line.}
\end{figure}
We note that, in the tests reported in this section, the preconditioner for each of the systems in Eqs.~\eqref{eqn:tempe_imp}--\eqref{eqn:vpari_imp} and \eqref{eqn:wave}--\eqref{eqn:diff} is recomputed whenever the number of iterations exceeds $20$.
Throughout this section, we consider two magnetic equilibrium configurations. The first equilibrium, reported in Fig.~\ref{fig:equil_ana}, is obtained by analytically solving the Biot–Savart law in the infinite aspect ratio limit, assuming a Gaussian current density profile within the simulation domain, and an external current filament located outside the domain to generate the X-point~\cite{giacomin2022turbulent}. The resulting poloidal magnetic flux function in the poloidal plane spanned by $(x,y)$ (see Fig.~\ref{fig:domain}) takes the form:
\begin{equation}
\begin{alignedat}{2}
    \Psi(x,y)=&\frac{1}{2}A_{m} \operatorname{Ei}\left[-\frac{((x-x_{m})^2+(y-y_{m1})^2)}{a_s^{2}}\right]-\frac{1}{2}A_{m}\log\left[(x-x_{m})^2+(y-y_{m1})^2\right]-\\&\frac{1}{2}A_{m}\log\left[(x-x_{m})^2+(y-y_{m2})^2\right];
    \label{eqn:psi}
\end{alignedat}
\end{equation}
where $\operatorname{Ei}(x)=\int_{- \infty}^x \frac{e^t}{t} \ dt$ and $A_{m}$, $x_{m}$, $y_{m1}$, $y_{m2}$ and $a_s$ are parameters reported in App.~\ref{sec:app_psi}. This configuration reproduces a magnetic topology of a tokamak with a lower single-null divertor, as illustrated in Fig.~\ref{fig:equil_ana}. Second, simulations based on the TCV-X21 experimental dataset~\cite{oliveira2022validation}, developed on the TCV tokamak for validation studies of turbulence codes, are performed. TCV-X21 is a lower single-null L-mode discharge carried out at low toroidal magnetic field. The computational domain used in our tests corresponds to, approximately, half the size of the TCV tokamak, see Fig.~\ref{fig:equil_sperim}.


Scalability tests requiring variations in the number of degrees of freedom are conducted using the analytical equilibrium, which allows for straightforward domain rescaling. The experimental equilibrium is instead used to assess the computational cost of the IMEX scheme in a realistic and relevant scenario.
All simulations reported in this section are initialized from a turbulent state.

\subsection{Verification and convergence}
\label{sec:verification}
The verification of the IMEX implementation in GBS is carried out using the method of manufactured solutions (MMS) \cite{riva2014verification}. The manufactured solutions $s_m$ for each unknown $m \in \{\theta, \Omega, \vpare, \vpari, t_e, t_i, \phi \}$ are chosen in the form
\begin{equation*}
s_m(y, x, z, t; \boldsymbol{\xi}_m) =
c_{0,m} \big[\alpha_m + \beta_m \sin\big(A_{y,m} y + \gamma\big)\big]
\sin\big(A_{z,m} z \big)
\sin\big(A_{x,m} x + A_{t,m} t\big),
\end{equation*}
where $\boldsymbol{\xi}_m = [c_{0,m}, \alpha_m, \beta_m, A_{y,m}, \gamma, A_{z,m}, A_{x,m}, A_{t,m}]^T $. 
 The parameters used in the tests discussed here are reported in App.~\ref{sec:app_vert}. All simulations used for the MMS verification employ the same spatial grid, with $N_x = N_y = N_z = 128$ and consider the analytical equilibrium given in Eq.~\eqref{eqn:psi}. 
 
 We note that the total error affecting the numerical solution results from the combined contribution of the discretization error and the iterative solver (tolerance) error, the latter being determined by the tolerance threshold prescribed by the user. We first focus on the time discretization error, by fixing the spatial discretization (i.e., keeping the grid unchanged), and perform tests with solver relative tolerances set to $10^{-10}$, to decouple the error associated with the iterative solvers from the discretization error.

\begin{figure}[t]
    \centering
    \includegraphics[width=1.05\linewidth]{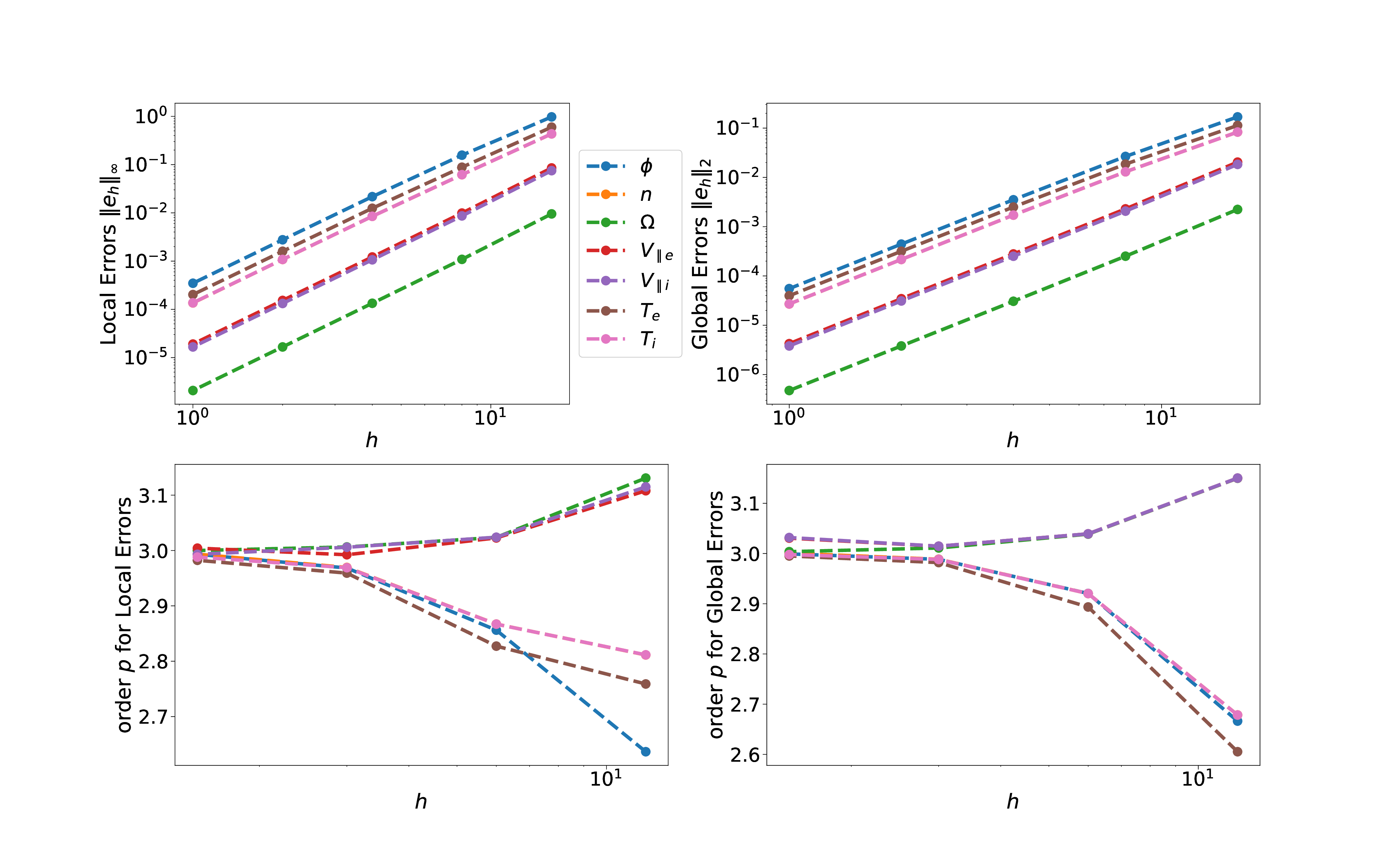}
    \caption{Local $l_{\infty}$ and global $l_{2}$ norms of the discretization error as a function of the time resolution parameter $h=\Delta t/ \Delta t_0$ where $\Delta t_0=1.25\times 10^{-5} $. Results are shown in dimensionless units.}
    \label{fig:convergence}
\end{figure}

In Fig.~\ref{fig:convergence}, we show the order of accuracy of the time integration method, estimated as
\begin{equation}
p = \frac{\log(e_{rh}/e_h)}{\log r},
\label{eqn:order_est}
\end{equation}
where $rh$ denotes a coarsening of the time step $h$ by a factor $r$, and $e_h=\|u_h-u\|$ represents the error norm (either $l_\infty$ or $l_2$) between the numerical solution $u_h$ obtained with time step $h$ and the analytical solution $u$.
The expected temporal order of convergence of the BPR(3,4,3) scheme is three, as confirmed by Fig.~\ref{fig:convergence}, which reports the values of $p$ computed according to Eq.~\eqref{eqn:order_est} using both the $l_\infty$ and $l_2$ norms.

\begin{figure}[t]
    \centering
    \includegraphics[width=1.05\linewidth]{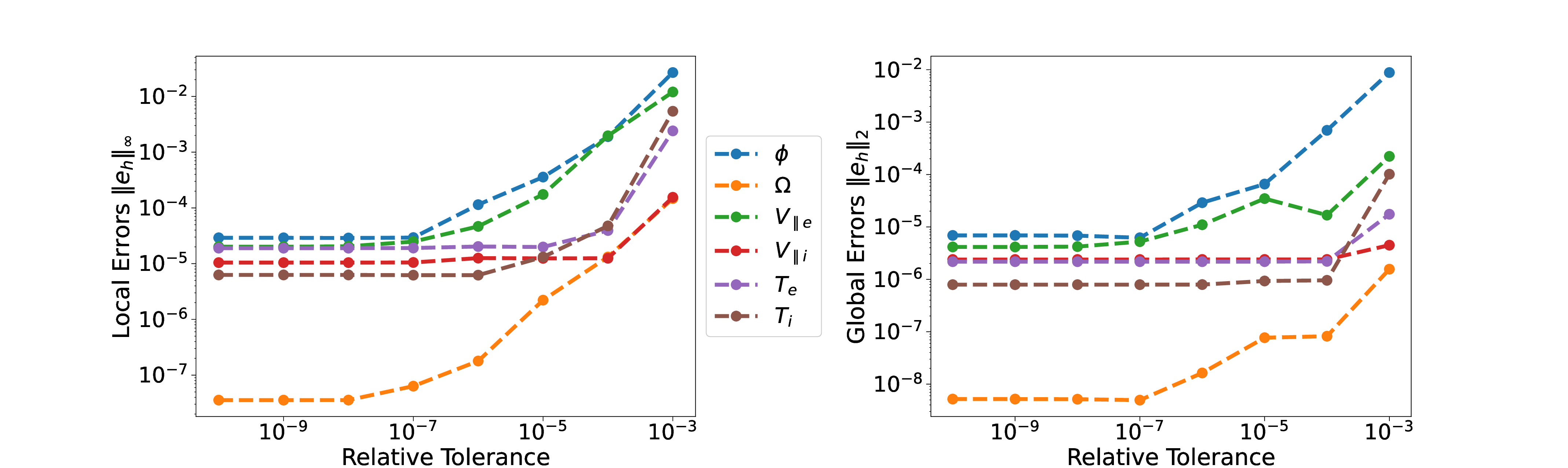}
    \caption{Local $l_{\infty}$ and global $l_{2}$ norms of the discretization error as a function of the relative tolerances set for the different linear systems. The time step is chosen to be $ \Delta t =10^{-3}$ and the grid size $N_x=160$, $N_y=212$, $N_z=40$. Results are shown in dimensionless units.}
    \label{fig:convergence_tol}
\end{figure}


To determine the impact of the convergence tolerance set for the iterative solvers used in the numerical experiments, a range of relative tolerances is tested and the corresponding error between the iterative solution and the manufactured solution is evaluated. This is reported in Fig.~\ref{fig:convergence_tol}. 
 The results show that the error decreases monotonically as the tolerance is reduced, to a value of approximately  $10^{-8}$. Below this threshold, the discretization error dominates and further tightening of the solver tolerance yields no appreciable improvement in accuracy. We note that this threshold is not universal: it depends on the condition number of the system matrix, which varies with the simulation state and the magnitude of the field gradients. 
 
In the typical setting of GBS simulations, where an exact solution is not available, and therefore the optimal tolerance cannot be determined from direct comparison against the reference solution, we identify the solver tolerance that is both necessary and sufficient to preserve the order of accuracy of the discretization, by estimating the order of accuracy through the Richardson extrapolation \cite{riva2014verification}, 
computed as
\begin{equation}
\hat{p} = \frac{\ln\big(\|u_{r^2h} - u_{rh}\|/\|u_{rh} - u_h\|\big)}{\ln(r)},
\label{eqn:phat}
\end{equation}
where $r$ denotes the grid refinement ratio, and comparing it with the theoretical value $p$. 

\begin{table}[t]
    \centering
  \begin{tabular}{|c|c|c|}
\hline
\multirow{2}{*}{Rel. Tol.} & \multicolumn{2}{|c|}{$\hat{p}$} \\
\cline{2-3}
&$<p_e>_{z,t}(y_0)$ &$<\phi>_{z,t}(y_0)$ \\
\hline
$10^{-10}$ & 2.2 & 2.7 \\
$10^{-8}$ & 2.2 & 2.7 \\
$10^{-6}$ & 2.1 & 2.7 \\
$10^{-4}$ & 1.7 & 1.8 \\
\hline
\end{tabular}
\vspace{2mm}
\caption{Values of $\hat{p}$ for the radial profiles of electron pressure and electrostatic potential, computed with different relative tolerances of the solvers. }
\label{tab:rde}
\end{table}
We highlight that, since the temporal discretization error is orders of magnitude smaller than the spatial discretization error, our analysis focuses on the latter. Consequently, computing $\hat{p}$ with Eq.~\eqref{eqn:phat} requires three simulations performed on grids of distinct spatial resolution: $80 \times 106\times 20$ (coarse) , $160 \times 212 \times 40 $ (medium), $320 \times 424 \times 80$ (fine), for each value of the relative tolerance. 
The estimated accuracy orders $\hat{p}$ are evaluated for the radial profiles of electron pressure and electrostatic potential, averaged over the toroidal direction and over one time unit, and their values are reported in Tab.~\ref{tab:rde}. 

For most configurations, $\hat{p}\approx2$, consistently with the expected theoretical value $p = 2$, a value due to the parallel Laplacian, whose second-order discretization dominates the asymptotic spatial error even though all other operators are fourth-order. The values of $\hat{p}$ slightly above $2$ obtained in some cases are attributable to the pre-asymptotic regime, in which the fourth-order operators raise the measured convergence rate above its asymptotic value. A notable exception is observed for simulations carried out at a relative tolerance of $10^{-4}$, where the order of accuracy deteriorates 
compared to results obtained at stricter tolerances, indicating that the iterative solver error is contaminating the solution. These findings indicate that, for the turbulent simulations considered here, a relative tolerance of $10^{-6}$ is sufficient to preserve the formal order of accuracy, with results comparable to those obtained at tighter tolerances. Nevertheless, $10^{-8}$ is selected for all subsequent simulations since, as shown in Fig.~\ref{fig:convergence_tol}, the error continues to decrease with tolerances set between $10^{-6}$ and $10^{-8}$ before plateauing. Choosing $10^{-8}$ ensures a robust margin against condition number variations that may arise in more demanding turbulent regimes.
\subsection{Parallel and algorithmic performance}
\label{sec:parall}
The linear systems in Eqs.~\eqref{eqn:imex}, \eqref{eqn:tempe_imp}--\eqref{eqn:vpari_imp}, \eqref{eqn:wave}--\eqref{eqn:diff} are solved using the PETSc library \cite{petsc-web-page}, with the parallel implementation relying on PETSc's Data Management for Structured Grids (DMDA) framework. The DMDA decomposes the computational domain into a Cartesian grid of rectangular parallelepipeds, each assigned a distinct MPI rank, matching the cubic structure induced by the discretization; the unknowns belonging to each sub-block are then mapped to the corresponding process. Data exchange between processes is performed through ghost-cell communication using MPI routines.


Strong scaling tests are performed by fixing the global problem size while increasing the number of processes, $n_p$, to evaluate the decrease of time-to-solution as $n_p$ grows.
On the other hand, weak scalability experiments are performed through successive mesh refinements that keep the number of grid points per process constant, as $n_p$ increases. 
In this section, we test the scalability of all linear systems individually, as well as the global scalability of the time stepping method.
The iteration counts and wall-clock times reported in the following sections are averaged over $30$ time steps, and the initial overhead associated with the preconditioner setup is not taken into account.
\subsubsection{Strong scaling}

The parallel speedup is defined as the ratio of the wall clock time (WCT) for running the software on a single process to the WCT for running the same software on $n_p$ processes, that is
\begin{equation*}
    \text{Speedup}= \frac{\text{WCT}(1)}{\text{WCT}(n_p)}
\end{equation*}

While, in the ideal case, the speedup is proportional to $n_p$, in practice, as the number of processes increases, the operations that do not scale with $n_p$, such as global communication and synchronization, cause the speedup to saturate and deviate from the ideal behavior.
Since the problem size exceeds the memory available on a single process for most configurations, the reference WCT used in the plots presented in this section is based on the smallest feasible number of processes rather than a single process.

\begin{figure}[!t]
    \centering
    \begin{subfigure}[b]{0.48\textwidth}
    \includegraphics[width=1.05\linewidth]{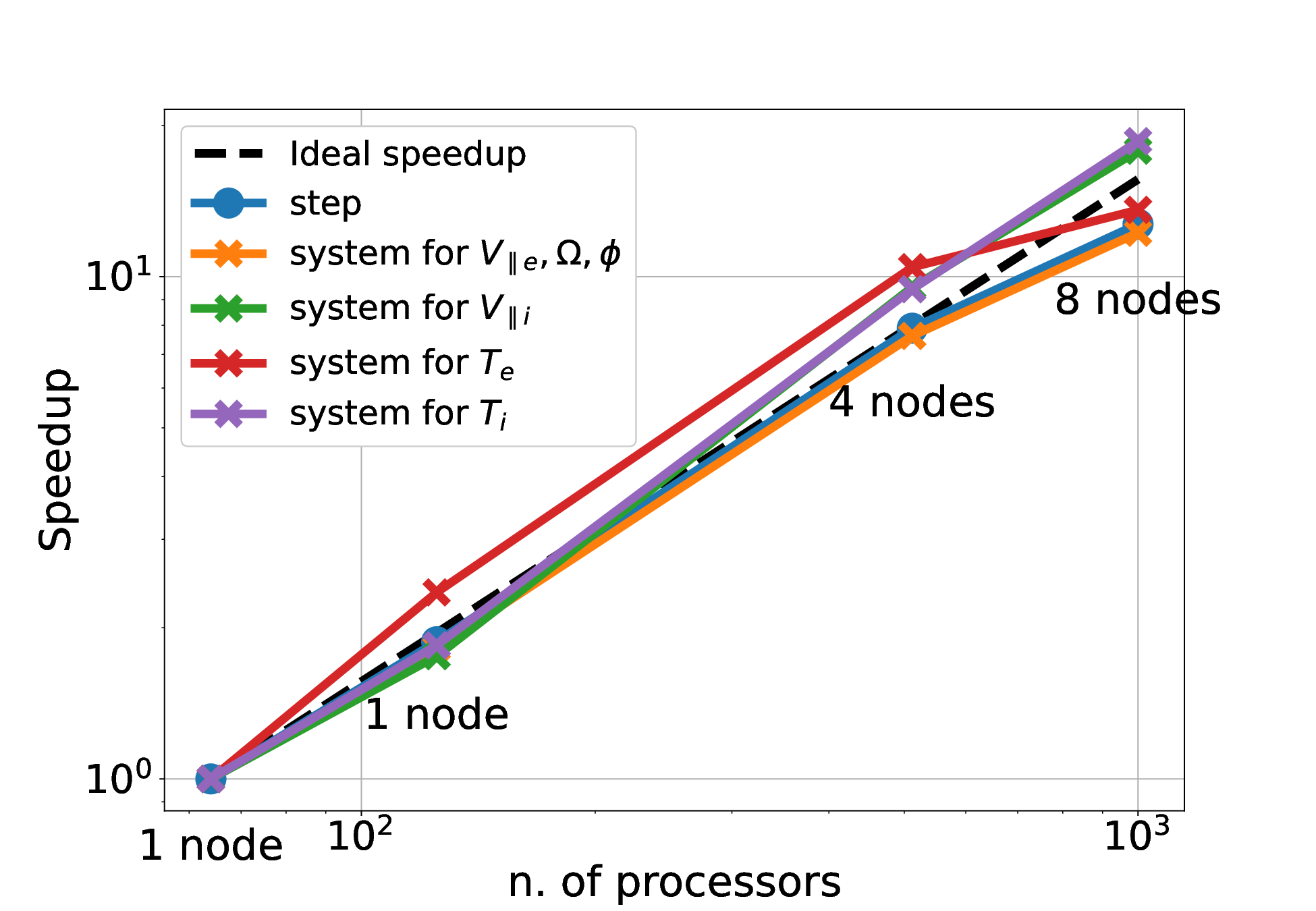}
    \caption{}
    \label{fig:strong_small}
     \end{subfigure}
      \begin{subfigure}[b]{0.48\textwidth}
         \includegraphics[width=\linewidth]{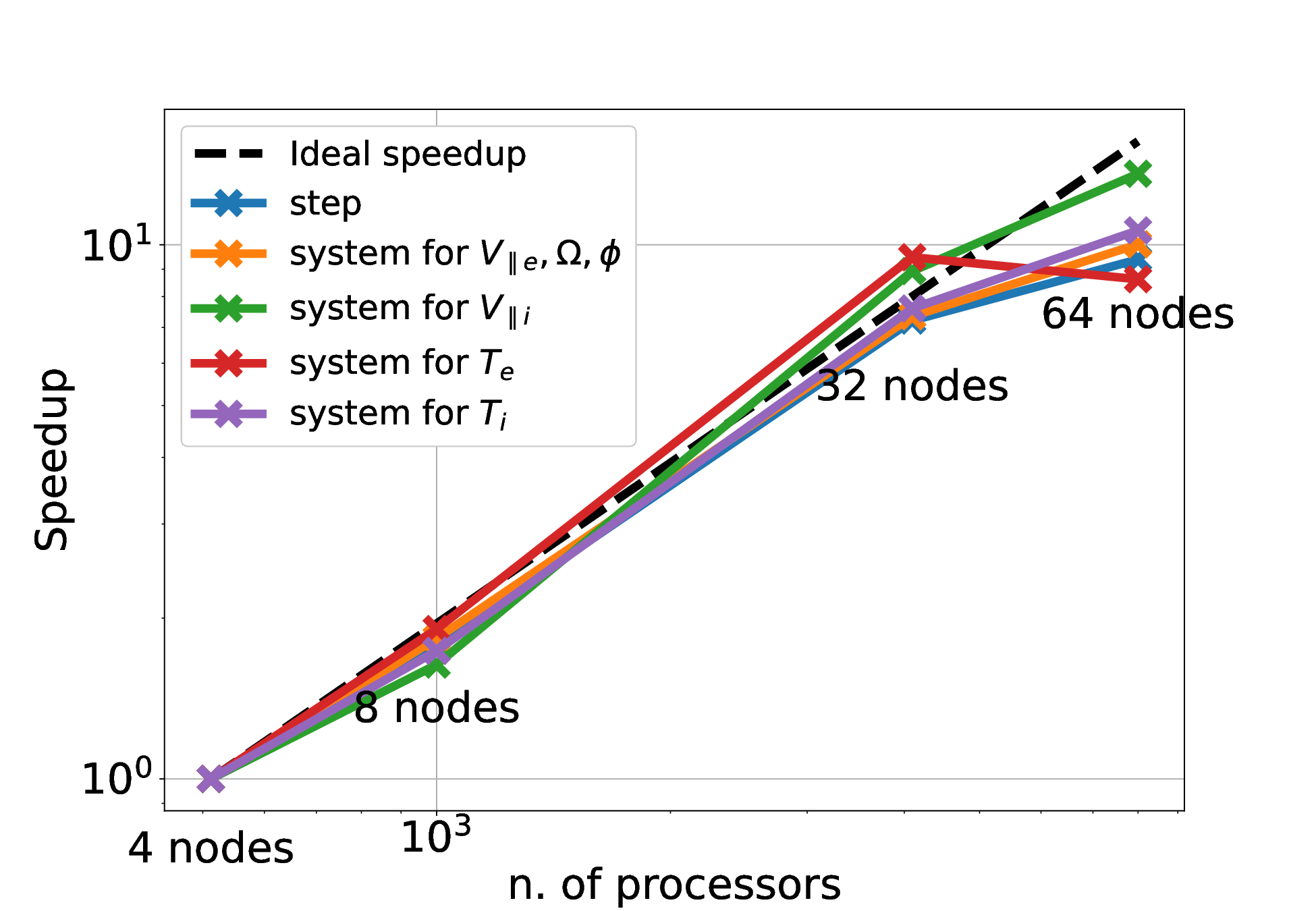}
    \caption{}
    \label{fig:strong_big}
     \end{subfigure}
     \caption{Strong scaling tests are performed using simulations with the analytical equilibrium for two different grid resolutions. In Fig.~\ref{fig:strong_small}, the grid size is $N_x = 160$, $N_y = 212$, and $N_z = 40$, with a time step $\Delta t = 10^{-3}$. In Fig.~\ref{fig:strong_big}, the resolution is increased to $N_x = 320$, $N_y = 424$, and $N_z = 80$, with a corresponding time step $\Delta t = 10^{-4}$.}
\end{figure}

Strong scaling is first assessed on a grid of size $N_x=160$, $N_y=212$, $N_z=40$, typical of simulations of a tokamak half the size of TCV, with processes distributed uniformly along the three spatial directions and their number increased simultaneously in each direction according to an $n\times n \times n$ decomposition, so that the total process count scales as $n_p=n^3$. The resulting speedup is shown in Fig.~\ref{fig:strong_small}, and the corresponding number of iterations for the linear systems in Eqs.~\eqref{eqn:imex}, \eqref{eqn:tempe_imp}--\eqref{eqn:vpari_imp}, \eqref{eqn:wave}--\eqref{eqn:diff} are reported in Tab.~\ref{tab:strong_scaling_small} as a function of the number of processes. The same analysis is then repeated on a more refined grid of size $N_x=320$, $N_y=424$, $N_z=80$, with the speedup and iteration counts reported in Fig.~\ref{fig:strong_big} and Tab.~\ref{tab:strong_scaling_big}, respectively.
\begin{table}[!t]
    \centering
    \begin{tabular}{|c|c|c|c|c|c|c|}
    \hline
        $n$ & iter. $(\Omega, \vpare, \phi)$ & iter. SAWs & iter. $\vpare$ & iter. $\vpari$  & iter. $T_e $ & iter. $T_i$\\
        \hline
            4&   6.73  &  1.59 &  1   &  4.5     &  6.8 &   2.5  \\
            5 &  6.98   &   1.63   &   1.11  &  4.74  &   5.46 &  2.53  \\
            8 &  7.04    &   1.73  &   1.17   &  4.84  &   6.85 &  2.53 \\
            10  &  7.04    &   1.87   &   1.11   &  4.75  &   9.92 &  2.53 \\
             \hline
    \end{tabular}
    \vspace{2mm}
    \caption{Average number of iterations for the simulations whose speedup is reported in Fig.~\ref{fig:strong_small}. The distribution of the processes along the three spatial directions is  equal to $n \times n \times n$. Here, "iter.\ $(\Omega, \vpare, \phi)$" refers to the average number of iterations to solve Eq.~\eqref{eqn:imex}, "iter.\ SAWs" to Eq.~\eqref{eqn:wave}, "iter.\ $\vpare$" to Eq.~\eqref{eqn:diff}, "iter.\ $\vpari$" to Eq.~\eqref{eqn:vpari_imp}, "iter.\ $T_e$" to Eq.~\eqref{eqn:tempe_imp}, and "iter.\ $T_i$" to Eq.~\eqref{eqn:tempi_imp}.}
    \label{tab:strong_scaling_small}
\end{table}
\begin{table}[!t]
    \centering
    \begin{tabular}{|c|c|c|c|c|c|c|}
    \hline
        $n$ & iter. $(\Omega, \vpare, \phi)$ & iter. SAWs & iter. $\vpare$ & iter. $\vpari$  & iter. $T_e $ & iter. $T_i$\\
        \hline
            8 &  4.74   &  1.66 &   1   &  2.96 &   3.69 &  1.69 \\
            10  &  5  &   1.54  &   1  & 3.29  &   3.67 &  1.69\\
            16  &  4.95  &   1.64  &   1  & 3.24  &   3.7 &  1.69 \\
            20  &  5.03 &   1.65  &  1   & 3.23 &  6.26 & 1.69  \\
             \hline
    \end{tabular}
    \vspace{2mm}
    \caption{Average number of iterations for the simulations whose speedup is reported in Fig.~\ref{fig:strong_big}. The distribution of the processes along the three spatial directions is equal to $ n\times n \times n$. Column labels are as in Tab.~\ref{tab:strong_scaling_small}.}
    \label{tab:strong_scaling_big}
\end{table}
In both cases, the results demonstrate good strong scaling behavior, up to 4 nodes for the smaller grid and up to 32 nodes for the larger grid. This trend is observed both in the wall-clock time per time step of the numerical scheme and in the time required to solve the system of Eqs.~\eqref{eqn:imex}. Furthermore, the number of iterations necessary to reach convergence remains approximately constant as the number of processes increases, indicating that the increased inter-process communications do not significantly affect the overall convergence properties of the solver.

\begin{figure}[!t]
    \centering
    \begin{subfigure}[b]{0.49\textwidth}
    \includegraphics[width=1.05\linewidth]{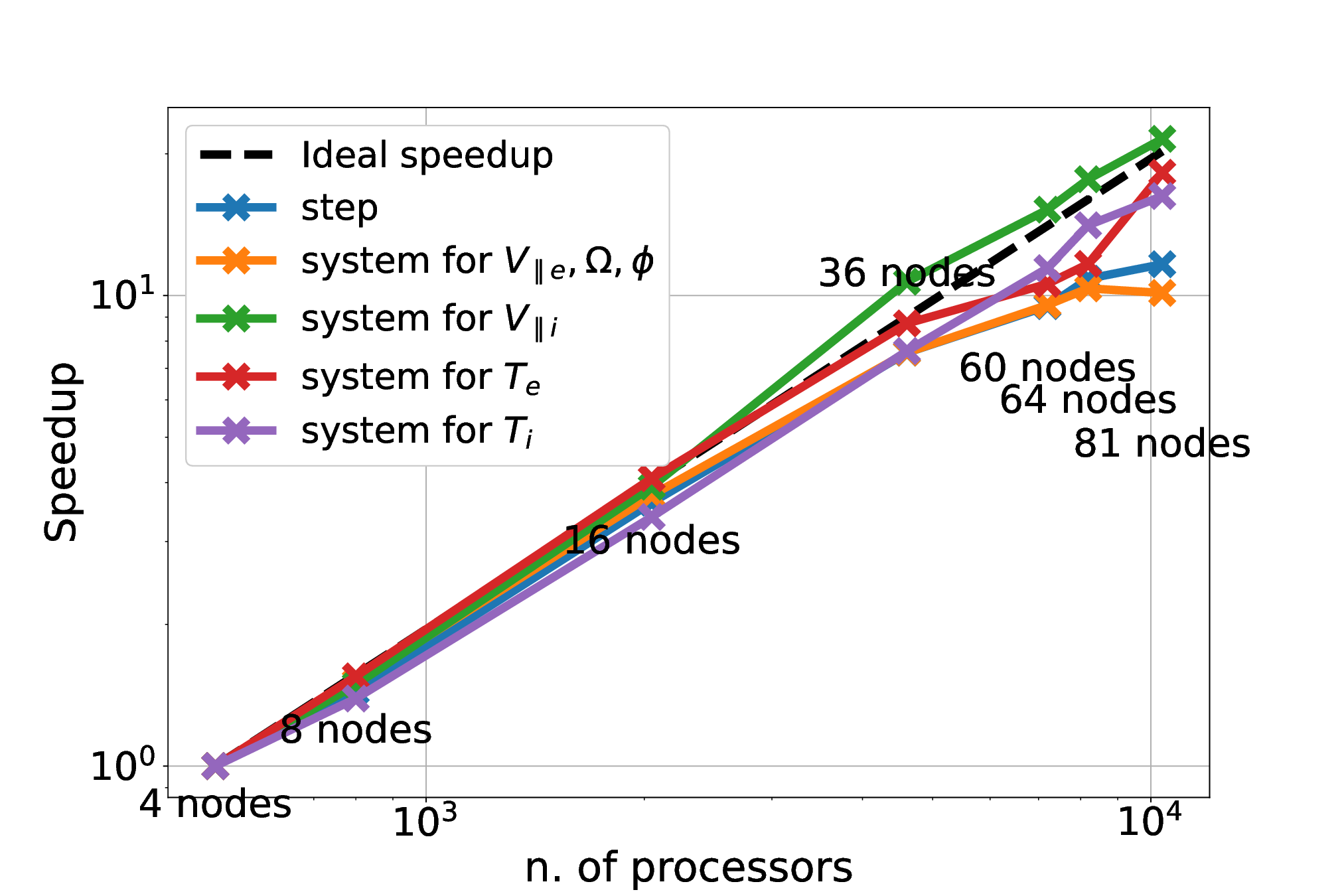}
    \caption{}
    \label{fig:strong_scaling_xy}
    \end{subfigure}
      \begin{subfigure}[b]{0.49\textwidth}
     \includegraphics[width=1.05\linewidth]{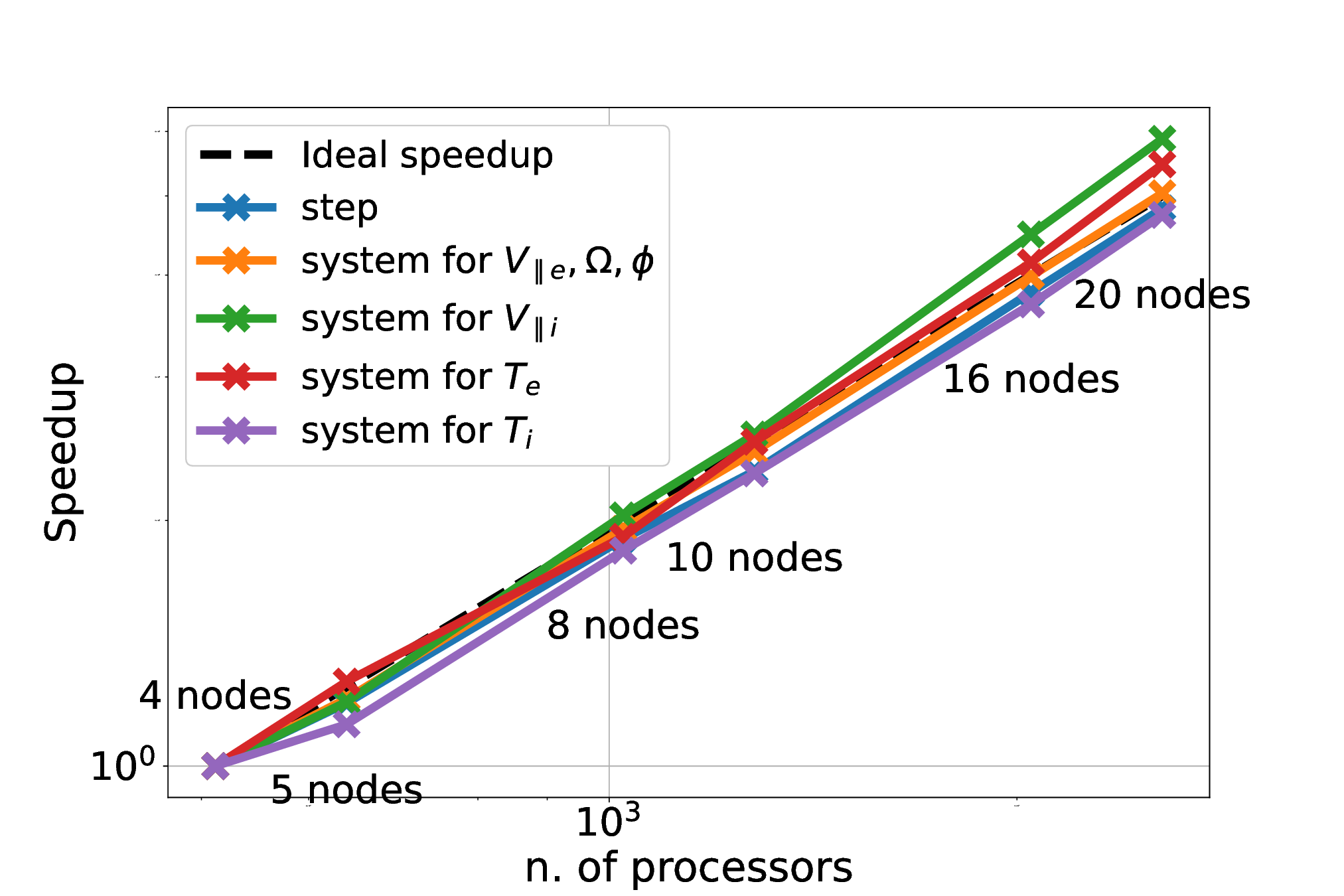}
    \caption{}
    \label{fig:strong_scaling_z}
    \end{subfigure}
    \caption{Strong scaling with analytical equilibrium in the direction $x$ and $y$ in Fig.~\ref{fig:strong_scaling_xy} and in $z$ in Fig.~\ref{fig:strong_scaling_z} with $\Delta t =10^{-4}$ and grid size $N_x=320$, $N_y=424$, $N_z=80$.}
\end{figure}
\begin{table}[t!]
    \centering
    \begin{tabular}{|c|c|c|c|c|c|c|}
    \hline
        $n$ & iter. $(\Omega, \vpare, \phi)$ & iter. SAWs & iter. $\vpare$ & iter. $\vpari$  & iter. $T_e $ & iter. $T_i$\\
        \hline
            8 &  4.74  &  1.66 &   1 &  2.96 &   3.69 &  1.68  \\
            10  &  4.82 &   1.6  &   1  & 2.85  &   3.65 &  1.68 \\
            16  &  4.77   &  1.65  &   1   & 2.88  &   3.66 &  1.68 \\
            24  &  4.82  &   1.8 &   1  & 2.92 &   4.75 &  1.69\\
            30  &  4.82  &   1.83 &   1  & 2.98 &   5.67 &  1.7\\
            32 & 4.82  & 1.85 & 1 & 2.91 & 4.82 & 1.69 \\
            36 &   4.8     &   1.89     &   1    &  2.94  & 4.11  & 1.69\\
             \hline
    \end{tabular}
    \vspace{2mm}
    \caption{Average number of iterations for the simulations whose speedup is reported in Fig.~\ref{fig:strong_scaling_xy}. The distribution of the processes along the three spatial directions is equal to $ n\times n \times 8$. Column labels are as in Tab.~\ref{tab:strong_scaling_small}.}
    \label{tab:strong_scaling_xy}
\end{table}
\begin{table}[!t]
    \centering
    \begin{tabular}{|c|c|c|c|c|c|c|}
    \hline
        $n$ & iter. $(\Omega, \vpare, \phi)$ & iter. SAWs & iter. $\vpare$ & iter. $\vpari$  & iter. $T_e $ & iter. $T_i$\\
        \hline
            4 &  4.78   &  1.64 &   1  &  2.91 &   3.398 &  1.69  \\
            5 &  4.77   &  1.64 &   1  &  2.94 &   3.4 &  1.69 \\
             8 &  4.78   &  1.64 &   1  &  2.9 &   3.63 &  1.69  \\
            10  &  4.77  &   1.64  &   1  & 2.9  &   3.64 &  1.69 \\
            16  &  4.77   &   1.64  &   1   & 2.91  &  3.65 &  1.7 \\
            20  &  4.77  &  1.64  &   1  & 2.88  &   3.66 &  1.7 \\
             \hline
    \end{tabular}
    \vspace{2mm}
    \caption{Average number of iterations for the simulations whose speedup is reported in Fig.~\ref{fig:strong_scaling_z}. The distribution of the processes along the three spatial directions is equal to $ 8\times 16 \times n$. Column labels are as in Tab.~\ref{tab:strong_scaling_small}.}
    \label{tab:strong_scaling_z}
\end{table}
To assess the scalability of the code under different parallelization strategies for process configuration, we perform two strong scaling tests using the largest grid, $320\times 424\times 80$. In the first test, processes are distributed as $n\times n\times 8$, where the number of processes in the $z$ direction is fixed, with the aim of evaluating scalability along the poloidal directions, $x$ and $y$. The corresponding speedup is reported in Fig.~\ref{fig:strong_scaling_xy}, and the number of iterations as a function of the number of processes is given in Tab.~\ref{tab:strong_scaling_xy}. In the second test, processes are distributed as $8\times 16\times n$, thus studying the scalability along the toroidal direction $z$. We note that, since each node contains 128 cores, increasing $n$ in this configuration simultaneously tests scalability across multiple nodes, making this also an inter-node scaling test. The speedup is reported in Fig.~\ref{fig:strong_scaling_z} and the iteration counts in Tab.~\ref{tab:strong_scaling_z}.
In both configurations, the code exhibits good scaling properties in all directions, and the number of solver iterations remains approximately constant as the number of processes increases, further confirming that inter-process communication does not degrade solver convergence.

\subsubsection{Weak scaling}

To analyze the weak scaling properties of our solvers, we examine both the algorithmic and the parallel (or implementation) scalability. Algorithmic scalability is evaluated by measuring the number of GMRES or FGMRES iterations required to achieve a chosen relative residual tolerance as the system size increases proportionally to the number of processes $n_p$. Parallel (or implementation) scalability is instead assessed by considering both the total wall-clock time (WCT) and the average cost per iteration (WCT/iteration). Together, these two analyses provide a comprehensive assessment of the implementation efficiency on distributed-memory architectures.
It is important to note that, since the time step is kept constant throughout these tests, while the mesh is refined, the linear systems solved at each stage become increasingly stiff as $n_p$ grows. 

\begin{figure}[!t]
    \centering
     \begin{subfigure}[b]{0.48\textwidth}
          \hspace{-2mm}
    \includegraphics[width=\textwidth]{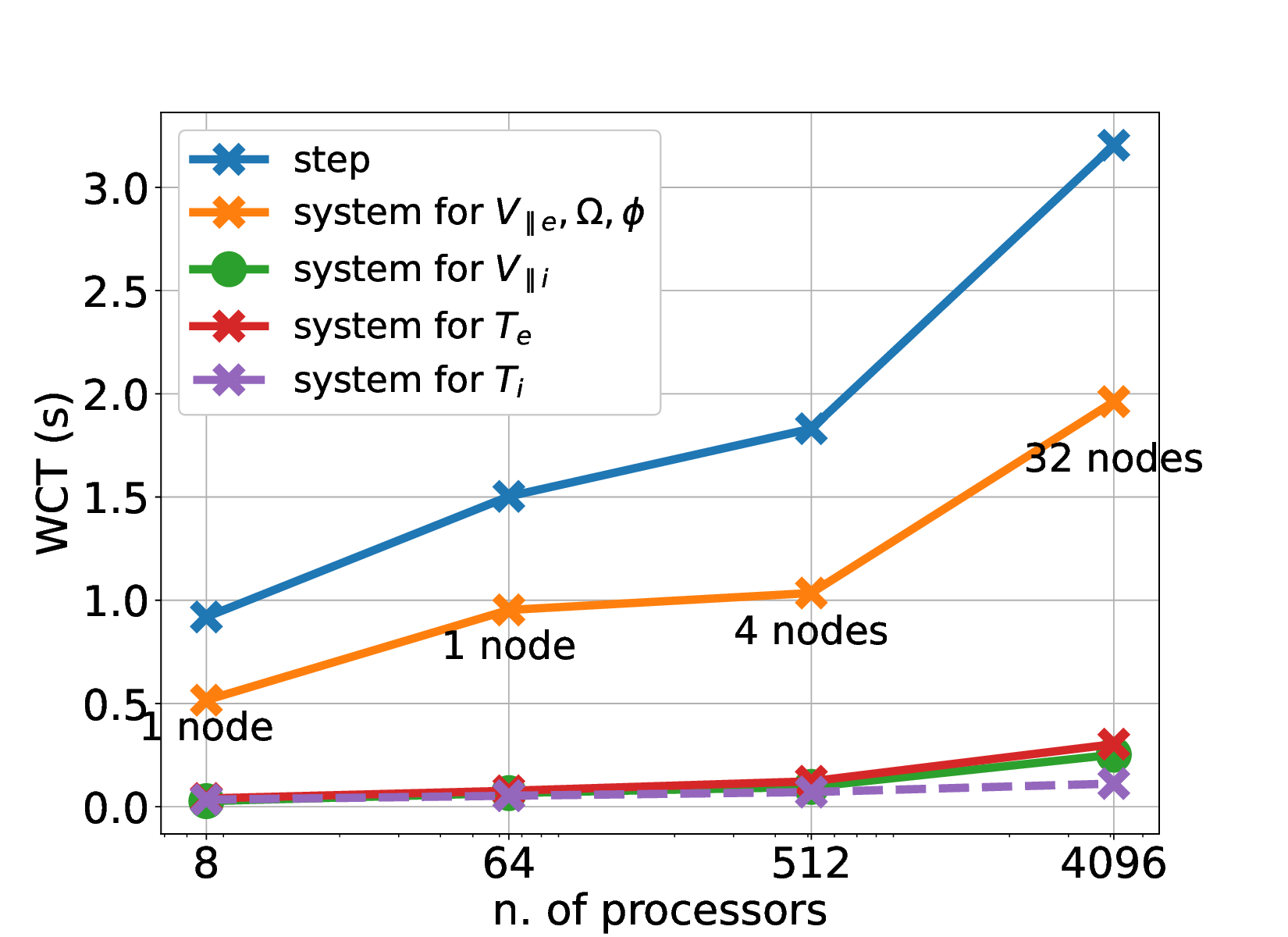}
    \caption{}
    \label{fig:weak_wct}
    \end{subfigure}
    \begin{subfigure}[b]{0.48\textwidth}
    \includegraphics[width=\textwidth]{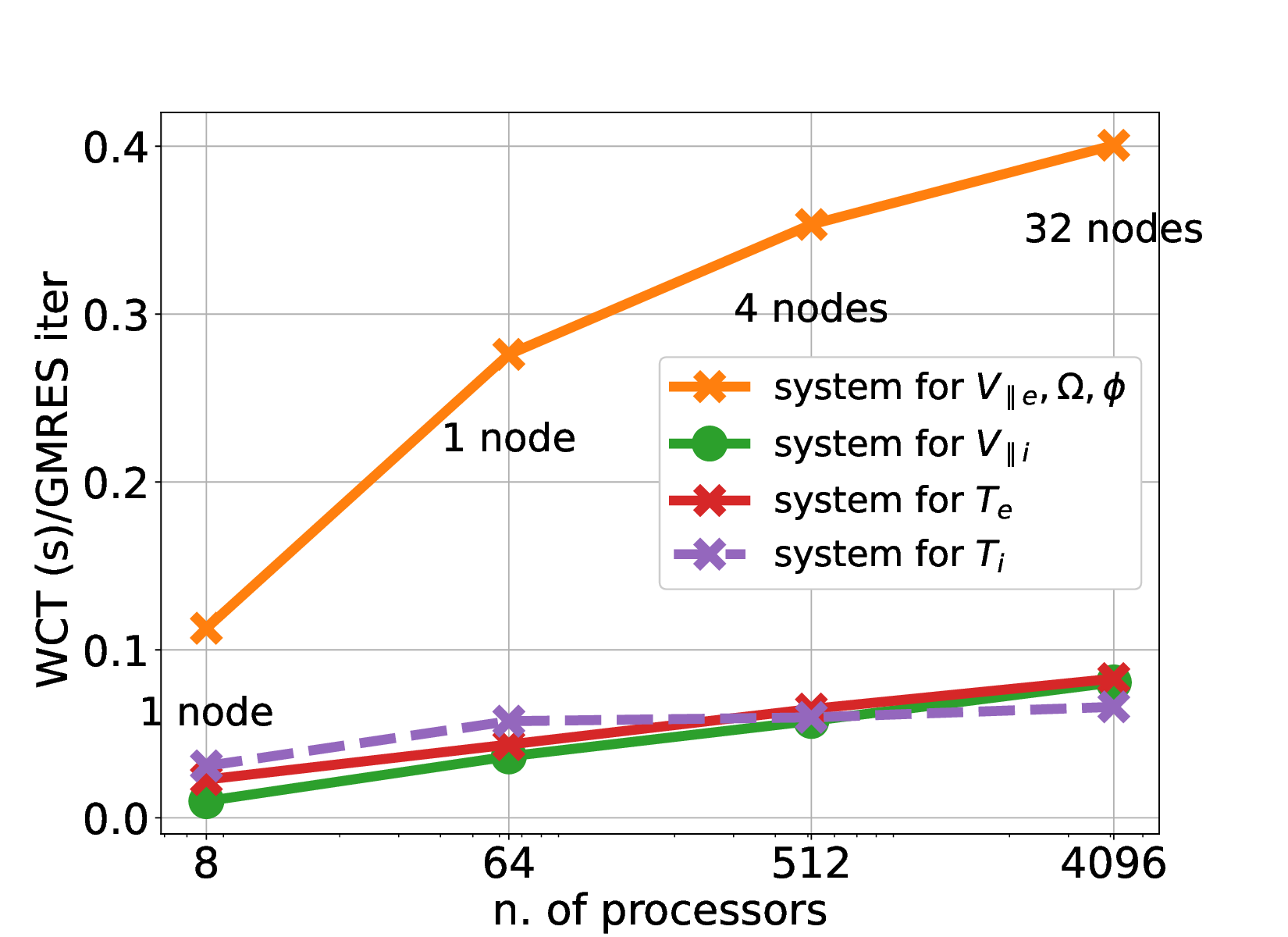}
    \caption{}
    \label{fig:weak_wct_iter}
    \end{subfigure}
    \caption{Weak scaling for simulations with analytical equilibrium and $\Delta t =10^{-4}$. Processes are distributed across the three spatial directions according to the $n\times n\times n$ decomposition. The wall-clock time (WCT) and the wall-clock time normalized by the number of GMRES iterations (WCT/iter.) are reported in seconds, with the corresponding iteration counts per system given in Tab.~\ref{tab:weak_scaling}.}
    \label{fig:weak}
\end{figure}
\begin{table}[!t]
    \centering
    \begin{tabular}{|c|c|c|c|c|c|c|c|c|c|c|}
   \hline
     $n$&  $n_p$ & mesh  & iter. $(\Omega, \vpare, \phi)$ & iter. SAWs & iter. $\vpare$ & iter. $\vpari$  & iter. $T_e $ & iter. $T_i$  \\
        \hline
         2& 8    &   $40\times53\times10$   &  4.57  &  1.63 &    1  &  2.72 &  1.76 & 1.1         \\
          4& 64   &  $80\times106\times20$    &   3.45  & 1.55     &   1  &  1.80   &  1.76 & 0.92               \\
         8& 512   & $160\times212\times40$ & 2.93 & 1.53 & 1 & 1.66 & 1.9 & 1.19  \\
           16& 4096  &  $320\times424\times80$ & 4.9  & 1.63 & 1 & 3.12  & 3.66 & 1.69 \\
             \hline
    \end{tabular}
    \vspace{2mm}
    \caption{Average number of iterations for the simulations in Fig.~\ref{fig:weak}. Processes are distributed across the three spatial directions according to the $n\times n\times n$ decomposition. Here, "iter.\ $(\Omega, \vpare, \phi)$" refers to the average number of iterations to solve Eq.~\eqref{eqn:imex}, "iter.\ SAWs" to Eq.~\eqref{eqn:wave}, "iter.\ $\vpare$" to Eq.~\eqref{eqn:diff}, "iter.\ $\vpari$" to Eq.~\eqref{eqn:vpari_imp}, "iter.\ $T_e$" to Eq.~\eqref{eqn:tempe_imp}, and "iter.\ $T_i$" to Eq.~\eqref{eqn:tempi_imp}.}
    \label{tab:weak_scaling}
\end{table}

Weak scaling is first assessed with processes distributed uniformly along the three spatial directions, such that the total process count is $n\times n\times n$. The resulting WCT are reported in Fig.~\ref{fig:weak_wct}, and the WCT of each linear system normalized to its number of iterations is reported in Fig.~\ref{fig:weak_wct_iter}. The number of iterations for each linear system, together with the corresponding grid size, is reported in Tab.~\ref{tab:weak_scaling}. For the system in Eq.~\eqref{eqn:imex}, since the number of internal iterations within the physics-based preconditioner remains approximately constant, as reflected by the number of iterations to solve Eq.~\eqref{eqn:wave} (SAWs equation) and the number of iterations to solve Eq.~\eqref{eqn:diff} ($\vpare$ equation) (see Tab.~\ref{tab:weak_scaling}), the WCT is normalized by the corresponding number of outer FGMRES iterations (iter. $(\Omega, \vpare, \phi)$).

\begin{figure}[!t]
    \centering
     \begin{subfigure}[b]{0.48\textwidth}
          \hspace{-2mm}
    \includegraphics[width=\textwidth]{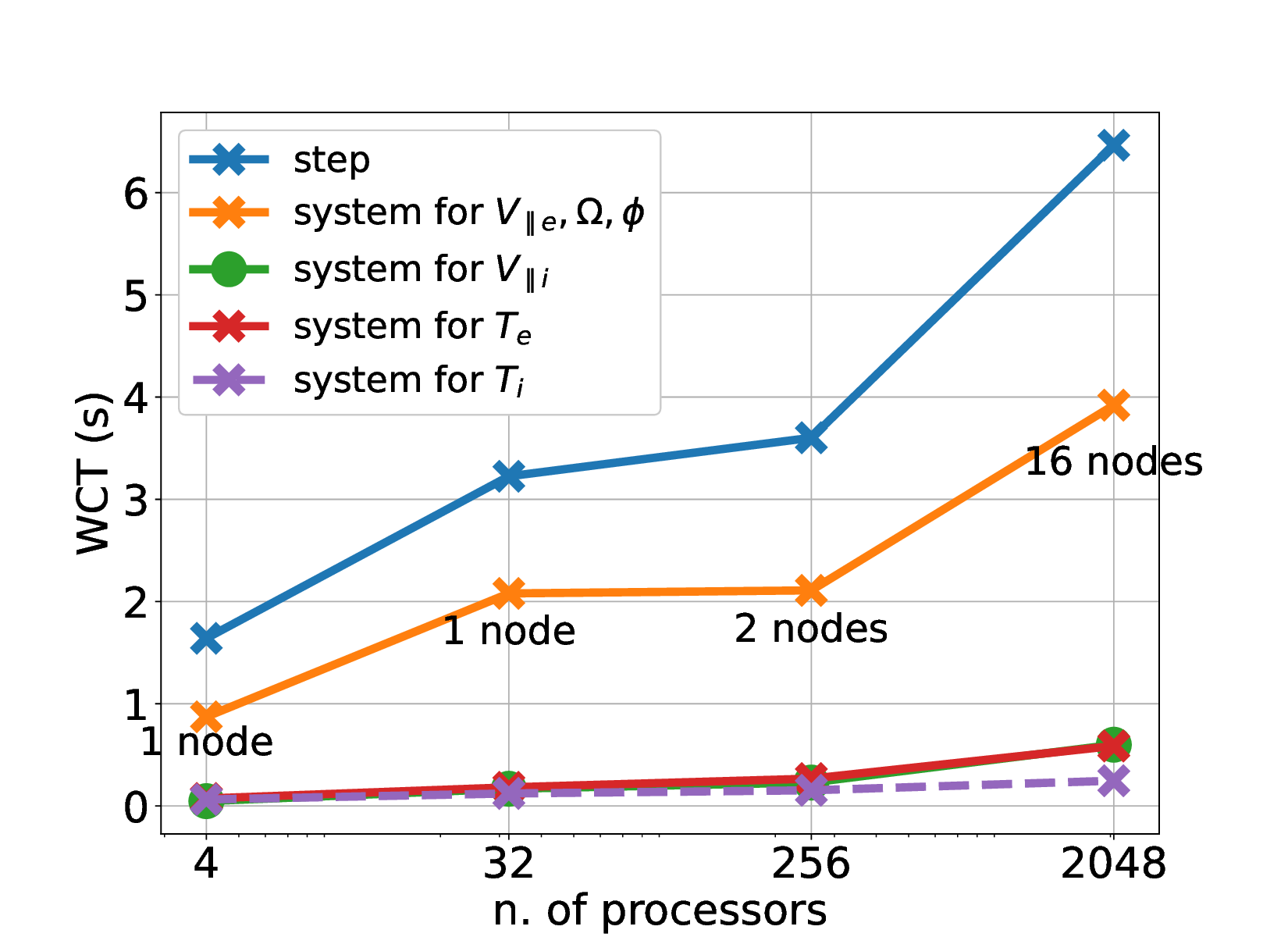}
    \caption{}
    \label{fig:weak_wct_nodes}
    \end{subfigure}
    \begin{subfigure}[b]{0.48\textwidth}
   \includegraphics[width=\textwidth]{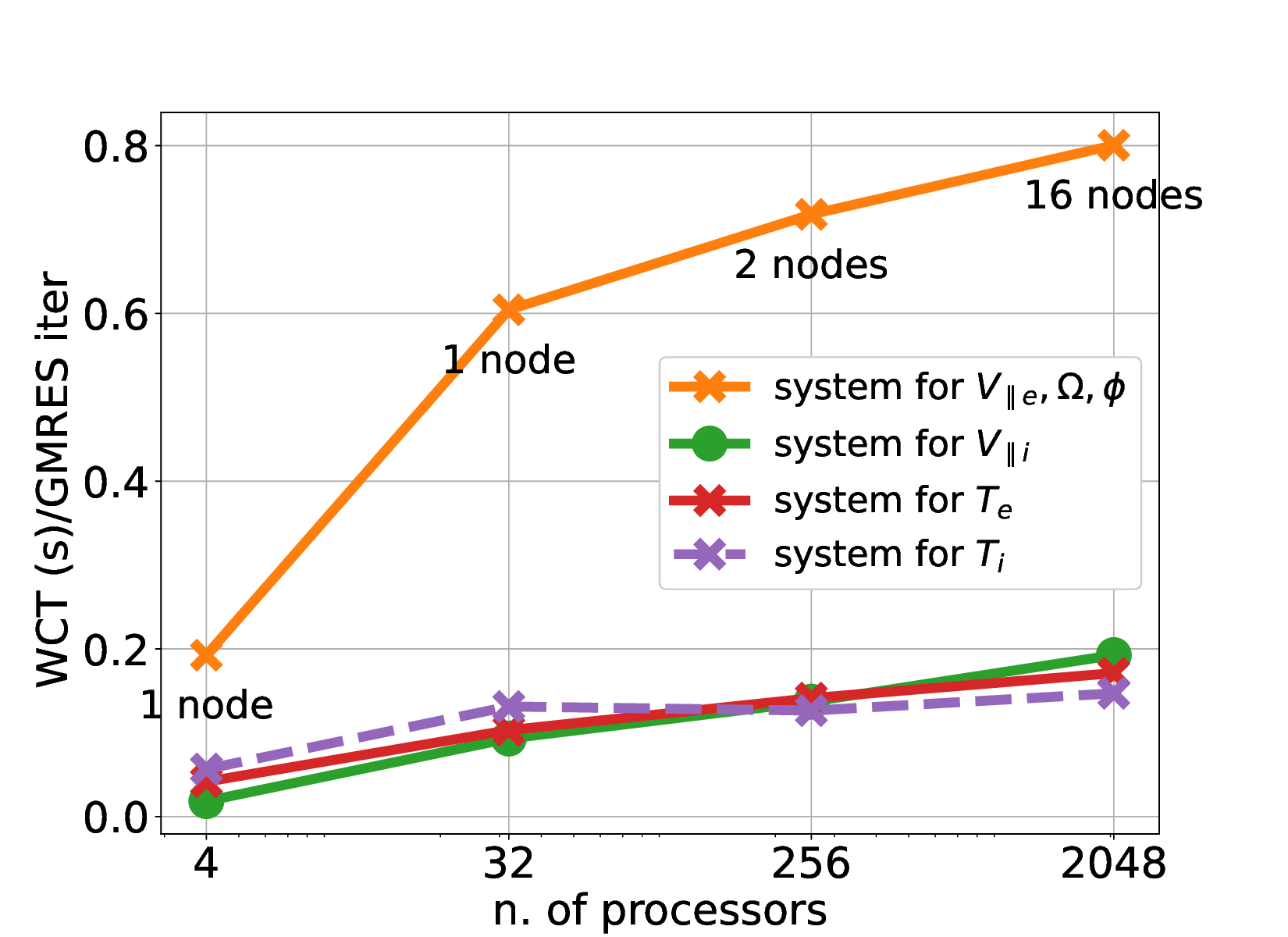}
    \caption{}
    \label{fig:weak_wct_iter_nodes}
    \end{subfigure}
    \caption{Weak scaling for simulations with analytical equilibrium and $\Delta t =10^{-4}$. Processes are distributed across the three spatial directions according to the $2n\times 2n\times n$ decomposition. The wall-clock time (WCT) and the wall-clock time normalized by the number of GMRES iterations (WCT/iter.) are reported in seconds, with the corresponding iteration counts per system given in Tab.~\ref{tab:weak_scaling_nodes}.}
    \label{fig:weak2}
\end{figure}
\begin{table}[!t]
    \centering
    \begin{tabular}{|c|c|c|c|c|c|c|c|c|c|c|}
    \hline
     $n$&  $n_p$ & mesh  & iter. $(\Omega, \vpare, \phi)$ & iter. SAWs & iter. $\vpare$ & iter. $\vpari$  & iter. $T_e $ & iter. $T_i$  \\
        \hline
         1& 4    &   $40\times53\times10$   &  4.53  &  1.74  &    1  &  2.54 &  1.75 & 1.11          \\
          2& 32   &  $80\times106\times20$    &   3.44  & 1.56   &   1  &  1.8  &  1.76 & 0.93             \\
         4&  256  & $160\times212\times40$ & 2.94 & 1.53 & 1 & 1.67 &  1.89 & 1.21  \\
          8 & 2048 &  $320\times424\times80$ & 4.9  & 1.63 & 1  & 3.11  & 3.43 & 1.68\\
             \hline
   \end{tabular}
       \vspace{2mm}
    \caption{Average number of iterations for the simulations in Fig.~\ref{fig:weak2}. Processes are distributed across the three spatial directions according to the $2n\times 2n\times n$ decomposition. Column labels are as in Tab.~\ref{tab:weak_scaling}.}
    \label{tab:weak_scaling_nodes}
\end{table}

 Weak scaling is subsequently assessed with an alternative process distribution of $2n\times 2n \times n$, which allocates more processes to the poloidal directions $x$ and $y$ than to the toroidal direction $z$. This choice reflects the typical grid structure of tokamak boundary simulations, in which the poloidal planes are resolved with a larger number of grid points than the toroidal direction. The resulting WCT are reported in Fig.~\ref{fig:weak_wct_nodes}, the WCT per iteration in Fig.~\ref{fig:weak_wct_iter_nodes}, and the iteration counts in Tab.~\ref{tab:weak_scaling_nodes}.

In both tests, the method maintains satisfactory \textit{algorithmic} scalability, with only a moderate increase in the number of iterations as the number of processes increases, as shown in Tabs.~\ref{tab:weak_scaling} and \ref{tab:weak_scaling_nodes}. Furthermore, as reported in Figs.~\ref{fig:weak_wct_iter} and \ref{fig:weak_wct_iter_nodes}, the WCT per iteration increases sublinearly with process count, exhibiting a $\log(n_p)$ behavior indicative of favorable \textit{parallel} scalability, consistent with the weak scaling results reported in \cite{CHACON2025113789}. 
Quantitatively, in the uniform $n\times n \times n$ distribution, the total WCT per time step increases by a factor of $3.5$ when the problem size is scaled by a factor of $512$. In the $2n\times 2n\times n$ distribution, the corresponding increase is a factor of $4$ for the same problem size scaling. These results confirm that the IMEX framework coupled with the physics-based preconditioner exhibits robust scalability across both process distribution strategies.
\subsection{Comparison of computational cost with explicit time stepping}
\label{sec:efficiency}

\begin{figure}[!t]
    \centering
   \includegraphics[width=0.85\linewidth]{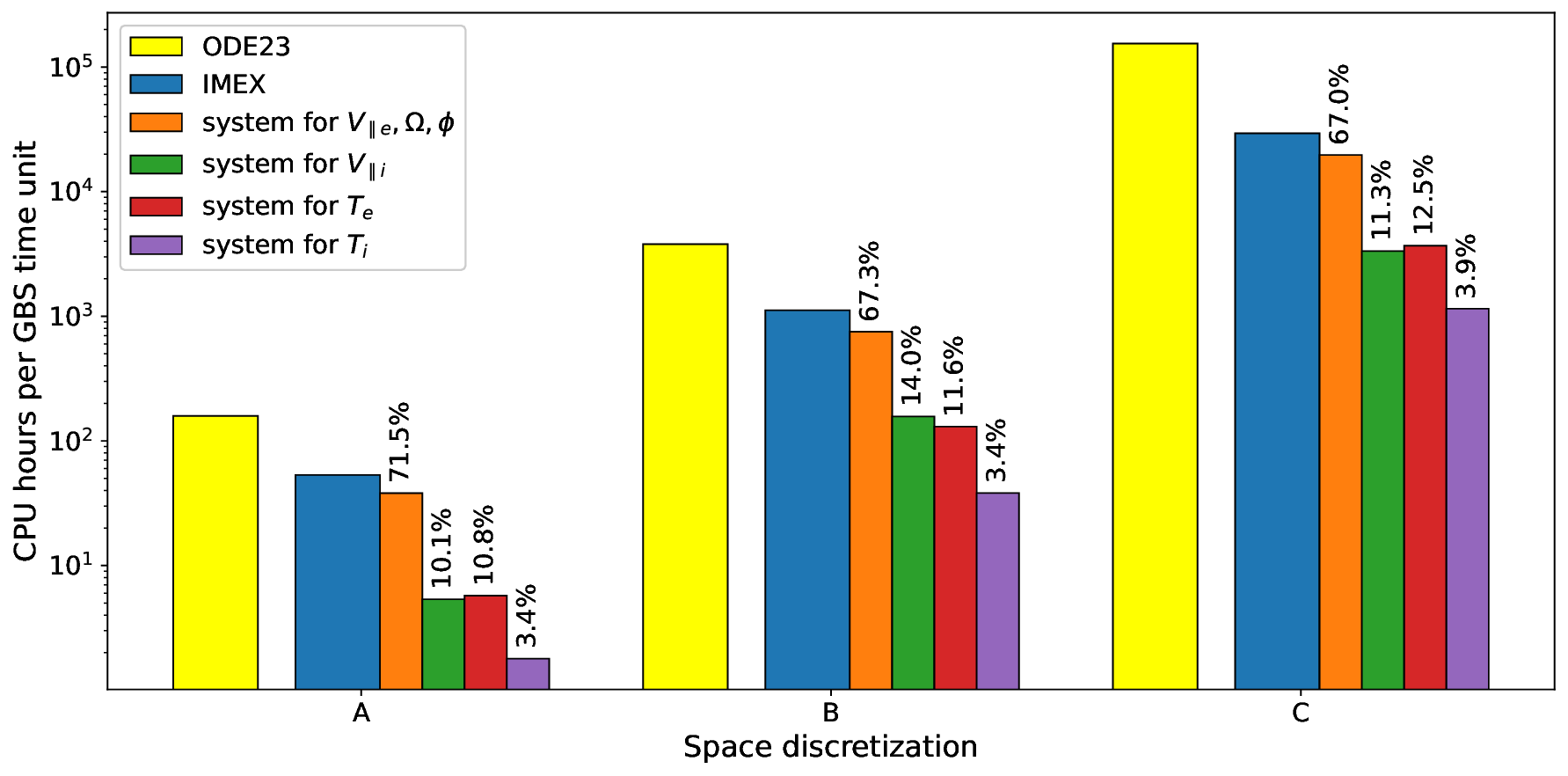}
    \caption{CPU hours per GBS time unit, measured over a simulation window of 3 GBS time units, displayed on a semi-logarithmic scale for three spatial resolutions and two time integration approaches: the adaptive explicit scheme ODE23 and the IMEX method equipped with the physics-based preconditioner introduced in this work. The total runtime of the IMEX scheme is split into the individual contributions arising from the different linear systems. The parameters used here are $m_i/m_e=200$ and $\nu=0.1$.}
    \label{fig:comp_eff}
\end{figure}

To quantify the computational cost of the IMEX scheme combined with the physics-based preconditioning strategy with respect to the explicit time stepping previously implemented, and to assess its dependence on the spatial discretization, we perform simulations on three progressively more refined grids: $80\times 106\times 20$ (grid A), $160\times 212\times 40$ (grid B), and $320\times 424\times 80$ (grid C). Starting from the same initial condition resulting from interpolation onto each grid, two series of simulations are carried out: the first employing the adaptive explicit time integration scheme ODE23 \cite{bogacki19893,shampine1997matlab}, and the second using the IMEX strategy described in Sec.~\ref{sec:imex}. The number of processes, increased with mesh refinement, is kept identical between the IMEX and ODE23 runs for each specific grid, thus ensuring a fair comparison. For both approaches, the time step is chosen as the largest value permitted by the stability constraints imposed by the remaining explicit terms in the equations. The computational efficiency of the two approaches is compared in terms of CPU hours per GBS time unit (equivalently, the wall-clock time per GBS time unit multiplied by the number of processes) as reported in Fig.~\ref{fig:comp_eff} and Tab.~\ref{tab:comp_eff}. This comparison is repeated for two sets of physical parameters, differing in the mass ratio $m_i/m_e$ and normalized Spitzer resistivity $\nu$: the first with $m_i/m_e=200$, $\nu=0.1$, and the second with $m_i/m_e=1200$, $\nu=0.05$, the larger mass ratio yielding a stiffer system. The remaining parameters used for these simulations are reported in App.~\ref{sec:app_vert}.

\begin{table}[!t]
    \centering
    \begin{tabular}{|c|c|c|c|c|c|c|c|}
    \hline
     mesh & nodes & tasks per node &  WCT ODE23 & WCT IMEX & Avg $\Delta t$ for ODE23 & $\Delta t$ for IMEX & \textbf{Speedup} \\

     \hline
    A & 1 & 64 & 8912& 2996& $6.1\times10^{-5}$ & $2\times10^{-3}$ &\textbf{2.97} \\
    B & 4 & 128 & 26640 &7860 & $1.65\times10^{-5}$& $10^{-3}$ & \textbf{3.39}   \\
   C & 16 & 128 & 271291 & 51664 & $3.8\times10^{-6}$ &$2\times10^{-4}$&\textbf{5.25} \\
    \hline
    \end{tabular}
    \vspace{2mm}
    \caption{Computational efficiency of the explicit (ODE23) and IMEX schemes, measured over 3 GBS time units, for grids A, B, and C, with $m_i/m_e = 200$ and $\nu = 0.1$. For each grid, the table reports the number of nodes and MPI tasks per node, the WCT in seconds per GBS time unit for each scheme, the average time step for the adaptive ODE23 scheme ("Avg $\Delta t$ for ODE23") and the fixed  time step for IMEX ("$\Delta t$ for IMEX"), and the resulting speedup, defined as the ratio of the two WCTs.}
    \label{tab:comp_eff}
\end{table}

In the test case with $m_i/m_e=200$, the results in Tab.~\ref{tab:comp_eff} demonstrate that the IMEX scheme consistently outperforms ODE23 across all grid sizes, achieving speedups of $2.97$, $3.39$, and $5.25$ on grids A, B, and C, respectively, with grid B representative of the spatial resolution typically employed in production simulations. The computational advantage of the IMEX approach therefore grows with mesh refinement.

Fig.~\ref{fig:comp_eff} reports the breakdown of the IMEX computational time into the contributions from the solution of each linear system. Notably, the relative cost associated with each linear system remains approximately constant as the number of degrees of freedom increases, indicating that the computational balance between the different solver components is preserved under mesh refinement. This is a positive indicator of the robustness of the preconditioning strategy: no single subsystem becomes a bottleneck as the problem size grows, and the computational load remains well balanced across the solvers. 
\begin{table}[!t]
    \centering
    \begin{tabular}{|c|c|c|c|c|c|c|c|c|}
    \hline
     mesh
     & \multicolumn{2}{c|}{syst. $(\Omega, \vpare, \phi)$} 
     & \multicolumn{2}{c|}{syst. $\vpari$} 
     & \multicolumn{2}{c|}{syst. $T_e$} 
     & \multicolumn{2}{c|}{syst. $T_i$} \\
  \cline{2-9}
     &  tot & solve/tot  & tot & solve/tot  & tot & solve/tot & tot & solve/tot \\
     \hline
    A & 2142 & 94.1\% & 301 & 94.6\% &  322& 90.8\% & 100 & 78.4\%\\
    B& 5289 & 94.7\% & 1104  & 96.2\% & 915 & 92.2\% & 268  & 80.7\% \\
   C &  34634 & 91.1\% & 5863 & 92.1\% & 6473 & 90.7\% & 2021 & 72.5\%\\
    \hline
    \end{tabular}
    \vspace{2mm}
   \caption{Breakdown of the WCT, in seconds per GBS time unit, for the IMEX scheme, measured over 3 GBS time units for grids A, B, and C. The contributions from the different linear systems solved at every time step are listed: the coupled SAW system $(\Omega, \vpare, \phi)$ [Eq.~\eqref{eqn:imex}], and the single-variable systems for $\vpari$, $T_e$, and $T_i$ [Eqs.~\eqref{eqn:vpari_imp}, \eqref{eqn:tempe_imp}, \eqref{eqn:tempi_imp}]. We consider $m_i/m_e = 200$ and $\nu = 0.1$. For each system, the total time (\textit{tot}) and its fraction spent in the linear solver (\textit{solve/tot}) are reported; the rest corresponds to the assembly/setup overhead.}
    \label{tab:comp_eff_imex}
\end{table}

The breakdown of the WCT for the IMEX scheme across each linear system being solved is reported in Tab.~\ref{tab:comp_eff_imex}. We notice that the fraction of time spent in the solve stage exceeds 90\% of the total time allocated to each linear system, except for the ion temperature system where, on average, only 77\% of the time is spent in the solve and the remaining fraction on matrix assembly and setup. This is likely because the ion parallel thermal conductivity is smaller than the electron one, making the ion temperature system less stiff than its electron counterpart and requiring fewer solver iterations to converge. 

\begin{figure}[!t]
    \centering
   \includegraphics[width=0.8\linewidth]{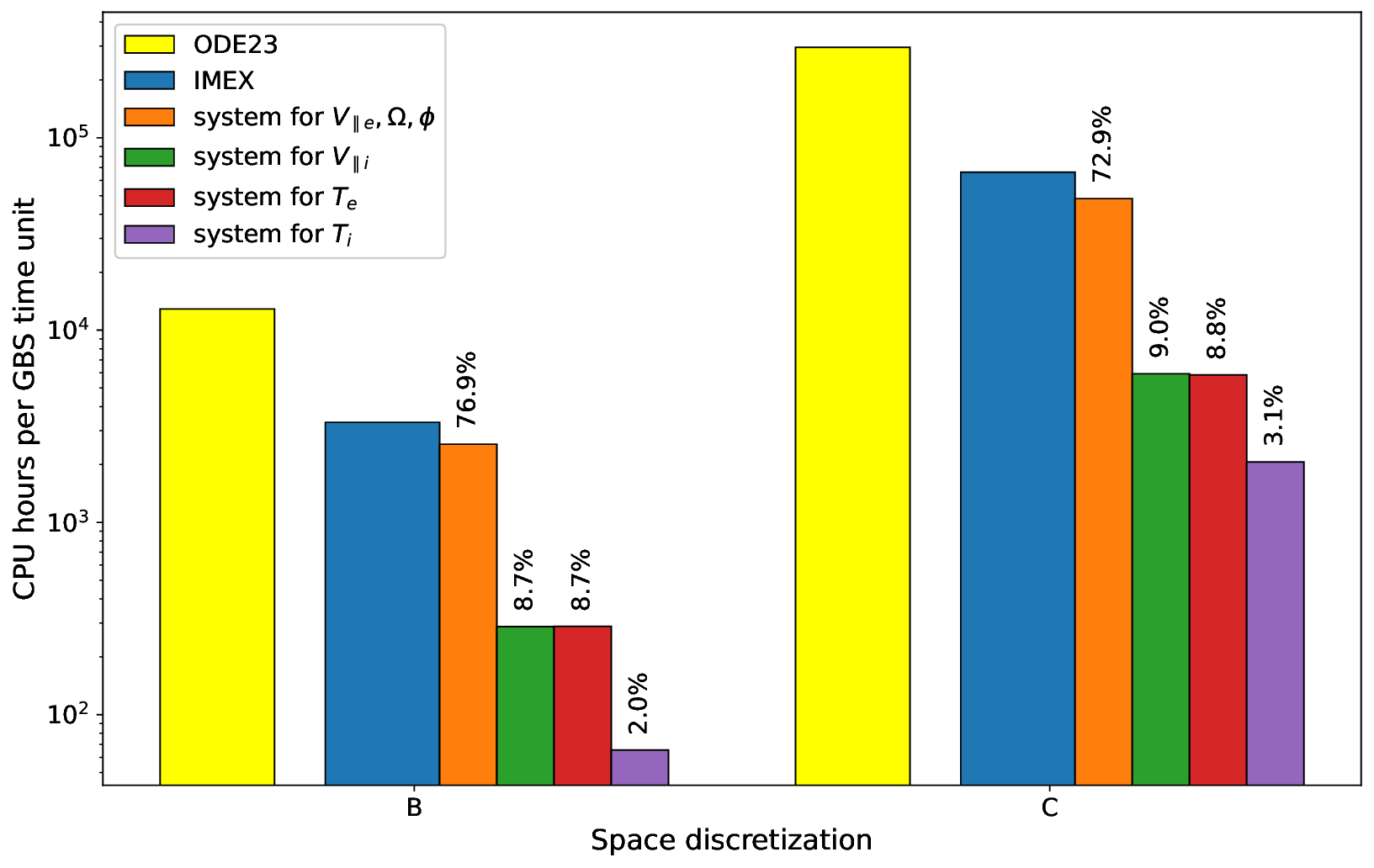}
   \caption{Same as Fig.~\ref{fig:comp_eff} for $m_i/m_e=1200$ and $\nu=0.05$.}
    \label{fig:comp_eff2}
\end{figure}

The test case with $m_i/m_e=1200$ and $\nu=0.05$ is representative of a regime in which the turbulent eddies have smaller sizes and simulations can be performed only with grids B and C. As in the first set of simulations, the computational cost per GBS time unit is much lower for IMEX than for ODE23 (see Fig.~\ref{fig:comp_eff2} and Tab.~\ref{tab:comp_eff2}), with speedups of $3.88$ and $4.46$ for grids B and C, respectively. Also in this case, the speedup increases with mesh refinement, but at a lower rate than that observed in the first set of simulations. We note that, as we increase the value of $m_i/m_e$ Eq.~\eqref{eqn:imex} becomes stiffer; indeed, the percentage of time spent solving it (orange columns in Fig.~\ref{fig:comp_eff2}) is larger than in the previous set of simulations (Fig.~\ref{fig:comp_eff}). At the same time, the overall system of equations becomes stiffer, as reflected by the decrease in the time step $\Delta t$ of both ODE23 and IMEX in Tab.~\ref{tab:comp_eff2} compared to the values in Tab.~\ref{tab:comp_eff}.
\begin{table}[!t]
    \centering
    \begin{tabular}{|c|c|c|c|c|c|c|c|}
    \hline

 \multirow{2}{*}{mesh} &  \multirow{2}{*}{nodes} & tasks  &  WCT & WCT & Avg $\Delta t$  & $\Delta t$ & \multirow{2}{*}{\textbf{Speedup}} \\
       & & per node & ODE23 &  IMEX & for ODE23 &  for IMEX & \\
     \hline
    B & 4 & 128 & 90539 & 23312 & $9\times10^{-6}$& $3\times10^{-4}$ & \textbf{3.88}   \\
   C & 16 & 128 & 519431 & 116449 & $1.9\times10^{-6}$ &$7\times10^{-5}$&\textbf{4.46} \\
    \hline
    \end{tabular}
    \vspace{2mm}
\caption{Same as Tab.~\ref{tab:comp_eff} for $m_i/m_e = 1200$ and $\nu = 0.05$.}
    \label{tab:comp_eff2}
\end{table}

\begin{table}[!t]
    \centering
    \begin{tabular}{|c|c|c|c|c|c|c|c|c|}
    \hline
     mesh
     & \multicolumn{2}{c|}{syst. $(\Omega, \vpare, \phi)$} 
     & \multicolumn{2}{c|}{syst. $\vpari$} 
     & \multicolumn{2}{c|}{syst. $T_e$} 
     & \multicolumn{2}{c|}{syst. $T_i$} \\
  \cline{2-9}
     &  tot & solve/tot  & tot & solve/tot  & tot & solve/tot & tot & solve/tot \\
     \hline
    B&  17930 & 95.1\% & 2019  & 93.8\% & 2022 & 90\% & 460 & 65.5\% \\
   C &  84918 & 90\% & 10422 & 88.8\% & 10271 & 83.9\% & 3622 & 60\%\\
    \hline
    \end{tabular}
    \vspace{2mm}
  \caption{Same as Tab.~\ref{tab:comp_eff_imex} for $m_i/m_e=1200$ and $\nu=0.05$.}
    \label{tab:comp_eff_imex2}
\end{table}
We finally remark, from Tab.~\ref{tab:comp_eff_imex2}, that the fraction of time spent in the solve stage per linear solver follows the same trend as in the set of simulations with lower $m_i/m_e$ (Tab.~\ref{tab:comp_eff_imex}); in particular, the ion temperature system exhibits the same behavior, with a proportionally larger share of the time spent on assembly and setup.
\subsection{Experimental Equilibrium}
\label{sec:exp}
In this section, we test the IMEX implementation in the case of an experimental equilibrium. Considering the magnetic flux $\Psi$ shown in Fig.~\ref{fig:equil_sperim}, result of the numerical resolution of Grad-Shafranov equation based on experimental measurements, we perform two tests, using different values of the coefficients that enter the ion and electron heat conduction in Eqs.~\eqref{eqn:flux_temp}: a first case with reduced heat conductivities and a second case with experimentally relevant values. 
The test allows us to estimate the speedup achievable with the IMEX scheme for an experimental equilibrium, and to assess how the increased stiffness associated with larger heat conductivities affects the computational time.
\begin{table}[!t]
    \centering
    \begin{tabular}{|c|c|c|c|c|c|c|c|c|c|c|}
    \hline
    \multirow{2}{*}{ $\Tilde{\chi}_{\parallel 0}^e$} &\multirow{2}{*}{ $\Tilde{\chi}_{\parallel 0}^i$} & WCT & WCT & WCT (\%) & WCT & WCT & WCT & Avg $\Delta t$ & $\Delta t$  & \multirow{2}{*}{ \textbf{Speedup} }\\
    & &  ODE23 & IMEX & $(\Omega, \vpare, \phi)$ & (\%) $\vpari$ & (\%) $T_e$ & (\%) $T_i$ & ODE23 & IMEX& \\
     \hline
    12 &10 & 219806  & 15740 & 62.7 & 7.4 & 12 & 5 &$3 \times 10^{-6}$ & $4\times10^{-4} $& \textbf{14}\\
    300 & 40 & 708456 & 20818 & 59 & 6 & 24.5 & 3.5 &$1\times 10^{-6}$ & $4\times10^{-4} $& \textbf{34}\\
    \hline
    \end{tabular}
    \vspace{2mm}
    \caption{Computational efficiency of the explicit (ODE23) and IMEX schemes for the experimental equilibrium, for the two cases considered: reduced ($\Tilde{\chi}_{\parallel 0}^e = 12$, $\Tilde{\chi}_{\parallel 0}^i = 10$) and experimental ($\Tilde{\chi}_{\parallel 0}^e = 300$, $\Tilde{\chi}_{\parallel 0}^i = 40$) parallel heat conductivities. For each case we report the WCT in seconds per GBS time unit of the two schemes; the percentage of the IMEX wall-clock time spent to solve the different systems [$(\Omega, \vpare, \phi)$, $\vpari$, $T_e$ and $T_i$]; the average time step $\Delta t$ used by each scheme; and the resulting speedup of the IMEX scheme over the explicit one.}
    \label{tab:comp_eff_exp}
\end{table}
 The grid size is $N_x = 140$, $N_y = 320$, $N_z = 64$; the remaining simulation parameters are reported in App.~\ref{sec:app_vert}. The fact that $\Psi$ is not a smooth function when an experimental equilibrium is used further increases the stiffness of the system, reducing the stability time step $\Delta t$ of the explicit time integration scheme.

As reported in Tab.~\ref{tab:comp_eff_exp}, temperature-dependent heat conductivities, especially with large normalized values, make the system stiff and impose a severe restriction on the time step of the explicit scheme, so that its overall computational cost becomes impractical at experimental values of the heat conductivities. The IMEX scheme, being free from this stability restriction, achieves a substantial speedup precisely in this regime, bringing the case of experimental conductivities within reach (second line of Tab.~\ref{tab:comp_eff_exp}). We further observe that the percentage of the IMEX computational time spent on the electron temperature solve grows appreciably from the reduced to the experimental conductivities, consistently with the increase in the electron conductivity. Overall, these results show that the IMEX scheme enables simulations with values of the Spitzer-Härm heat conductivities that are in the range observed in experiments, prohibitively expensive with an explicit integration.
\section{Conclusions}
\label{sec:concl}
In this work, a parallel, scalable Implicit-Explicit (IMEX) Runge–Kutta time integration scheme with a physics-based preconditioning strategy is developed for the drift-reduced Braginskii equations and implemented in the GBS code. The stiffest plasma dynamics, governed by SAWs and parallel diffusion, are treated implicitly using the globally stiffly accurate BPR(3,4,3) scheme, while the remaining non-stiff terms are advanced explicitly, enabling time steps well beyond the fully explicit CFL limit, while avoiding the cost and complexity of a fully implicit formulation. To efficiently solve the implicit subsystem, we construct a three-dimensional physics-based preconditioner inspired by the parabolization strategy of \citeauthor{chacon2002implicit}, extended here to account for both the SAW dynamics and the electron parallel viscosity.

The framework is verified through the method of manufactured solutions and demonstrates algorithmic and parallel scalability up to 8000 MPI tasks. A systematic comparison against the explicit scheme ODE23 shows a speedup of, approximately, $3.5$ on typical production grids, rising to, approximately, $5$ on the finest grid considered, depending on the input parameters, with the speedup growing with mesh refinement. 
Additionally, the implicit treatment of stiff parallel conduction terms allows the temperature dependence of the thermal conductivity coefficients to be incorporated for the first time within GBS, without incurring the prohibitive cost that such stiff terms would entail in a fully explicit formulation. These results show that the IMEX scheme enables simulations at experimental values of heat conductivities, prohibitively expensive with an explicit integration.

Several directions for future work are envisioned. First, the physics-based preconditioner strategy will be extended to the electromagnetic case, which might affect particularly the simulations of large size tokamaks. Indeed, the frequency of SAWs does not grow unboundedly with system size, when electromagnetic effects are accounted for. Second, we plan to investigate the porting of the code to GPU architectures, with the aim of preserving and potentially further improving the scalability properties demonstrated in the present work, thereby enabling high-fidelity simulations at the spatial and temporal scales required for reactor-relevant predictions.

\section*{Data Availability Statement}
The data that supports the findings of this study is available from the corresponding author upon reasonable request.
\section*{Acknowledgements}
The authors would like to thank L. Chacón for the helpful discussions about physics-based preconditioners.
 The simulations presented here were carried out in part at the Swiss National Supercomputing Centre (CSCS) under the Projects ID lp29. We also acknowledge EuroHPC Joint Undertaking for awarding the project ID EHPC-EXT-2024E02-100 access to LUMI at CSC, Finland. This work has been carried out within the framework of the EUROfusion Consortium, partially funded by the European Union via the Euratom Research and Training Programme (Grant Agreement No 101052200 — EUROfusion). The Swiss contribution to this work has been funded by the Swiss State Secretariat for Education, Research and Innovation (SERI). Views and opinions expressed are however those of the author(s) only and do not necessarily reflect those of the European Union, the European Commission, or SERI. Neither the European Union nor the European Commission nor SERI can be held responsible for them.

\bibliographystyle{unsrtnat}
\newpage
\bibliography{references}  

\clearpage
\appendix
\section{Appendix}
\subsection{Definition of the thermal conductivities in GBS}
\label{sec:app_therm}
The parallel thermal conductivities are
\begin{subequations}
\begin{align}
    &\chi_{\parallel e}=3.16 \frac{n_e T_e \tau_e}{m_e}\\
    &\chi_{\parallel i}=3.9 \frac{n_i T_i \tau_i}{m_i}
\end{align}
\label{eqn:chi}%
\end{subequations}
where the electron and ion collision times are
\begin{subequations}
\begin{align}
    &\tau_e=\frac{3\sqrt{m_e}}{4\sqrt{2\pi}}\frac{(4\pi\epsilon_0)^2}{Z^2e^4}\frac{T_e^{3/2}}{n\lambda}\\
    &\tau_i= \frac{3\sqrt{m_i}}{4\sqrt{2\pi}}\frac{(4\pi\epsilon_0)^2}{Z^4e^4}\frac{T_i^{3/2}}{n\lambda}
\end{align}
\label{eqn:col_tim}%
\end{subequations}
as reported in \cite{giacomin2022turbulent}, with $\lambda$ the Coulomb logarithm.
Inserting Eqs.~\eqref{eqn:col_tim} into Eqs.~\eqref{eqn:chi} makes the $T_a^{5/2}$ dependence of the parallel thermal conductivities explicit, so that $\chi_{\parallel e}=\chi_{\parallel 0}^e T_e^{5/2}$ and $\chi_{\parallel i}=\chi_{\parallel 0}^i T_i^{5/2}$. Normalizing $\chi_{\parallel 0}^e$ and $\chi_{\parallel 0}^i$ yields the coefficients
\begin{subequations}
\begin{align}
    \Tilde{\chi}^e_0=& \frac{2.37}{\sqrt{2\pi}}\sqrt{\frac{m_i}{m_e}}\frac{(4\pi\epsilon_0)^2}{Ze^4}\frac{T_{e0}^2}{\lambda n_0R_0}\\
    \Tilde{\chi}^i_0=& \frac{2.925}{\sqrt{2\pi}}\frac{(4\pi\epsilon_0)^2}{Z^2e^4}\frac{T_{e0}^2\tau^{5/2}}{\lambda n_0 R_0}
\end{align}
\end{subequations}
which enter Eqs.~\eqref{eqn:flux_temp}.
For typical physical scenarios, $\Tilde{\chi}^e_0$ varies over a range of $400$--$600$ and $\Tilde{\chi}^i_0$ over $10$--$20$.
\subsection{Butcher tableaux of BPR(3,4,3)}
\label{sec:app_butcher}
An IMEX-RK scheme is entirely characterized by its double \textit{tableau}, in which the explicit part is displayed before the implicit part, and takes the general form:
\begin{multicols}{2}
\begin{center}
\begin{tabular}{ c| c  }
$c$ & $A$ \\
\hline\rule{0pt}{10pt} 
& $b^T$
\end{tabular}
\end{center}
\columnbreak
\begin{center}
\begin{tabular}{ c| c }
$\Bar{c}$ & $\Bar{A}$\\
\hline\rule{0pt}{10pt} 
 & $\Bar{b}^T$
\end{tabular}.
\end{center}
\end{multicols}

For BPR(3,4,3), the double \textit{tableau} reads:
\begin{multicols}{2}
\begin{center}
$(\textbf{A},\textbf{b},\textbf{c})$\\
\begin{tabular}{ c| c c c c c }
$0$ &$0$ & & & &\\
$1$ & $1$ & $0$ & & &\\
$2/3$ & $4/9$ &$2/9$ & $0$ & &\\
$1$ & $1/4$ &$0$ & $3/4$ & $0$ & \\
$1$ & $1/4$ &$0$ & $3/4$ & $0$ & $0$ \\
\hline
& $1/4$ &$0$ & $3/4$ & $0$ & $0$ 
\end{tabular}
\end{center}
\columnbreak
\begin{center}
$(\Bar{\textbf{A}},\Bar{\textbf{b}},\Bar{\textbf{c}})$\\
\begin{tabular}{ c| c c c c c }
$0$ &$0$ & & & &\\
$1$ & $1/2$ & $1/2$ & & &\\
$2/3$ & $5/18$ &$-1/9$ & $1/2$ & &\\
$1$ & $1/2$ &$0$ & $0$ & $1/2$ & \\
$1$ & $1/4$ &$0$ & $3/4$ & $-1/2$ & $1/2$ \\
\hline
&  $1/4$ &$0$ & $3/4$ & $-1/2$ & $1/2$ 
\end{tabular}
\end{center}
\end{multicols}

\subsection{Parameters in the analytical function of the equilibrium}
\label{sec:app_psi}
Here we report the parameters used to define $\Psi$ in Eq.~\eqref{eqn:psi}:
\begin{align*}
    A_{m}=&\frac{25L_x}{12}L_y\\
    x_{m}=&\frac{L_x}{2}\\
    y_{m1}=&\frac{5}{8}L_y\\
    y_{m2}=&18\left(\frac{L_y}{40}\right)-y_{m1}\\
    a_s=&\frac{5}{40}L_y&
\end{align*}
where $L_x$ and $L_y$ denote the domain extents in the $x$ and $y$ directions, respectively. Unless otherwise specified, we set $L_x = 300$ and $L_y = 400$ in all simulations.

\subsection{Parameters used in the scaling and numerical tests}
\label{sec:app_vert}
The verification tests, whose results are shown in Fig.~\ref{fig:convergence}, are performed with $\rho^{-1}_* = 100$, $\eta_{0e} = \eta_{0i} = 1$, $m_i/m_e = 1$, and $\nu = 1$. The numerical tests are performed on a computational domain of size $L_x = 37.5\rho_{s,0}$ in the radial direction and $L_y = 50\rho_{s,0}$ in the vertical direction where $\rho_{s0}$ is the reference ion sound Larmor radius. The coefficients $\Tilde{\chi}^e_{\parallel 0}$ and $\Tilde{\chi}^i_{\parallel 0}$ in the flux limiter for parallel conduction are both set to 1. Table~\ref{tab:verif_param} reports the parameters selected for the manufactured solutions for each unknown.
\begin{table}[h]
    \centering
    \begin{tabular}{|c|c|c|c|c|c|c|c|c|}
    \hline
        $m$ &$c_0$ & $\alpha$ & $\beta$ &  $\gamma$ & $A_y$ & $A_x$ & $A_{z}$& $A_t$ \\
        \hline
           $\theta$ &  $1$     &    $1.1$  &   $0.4$   &  $0.2$       &    $\frac{2\pi}{L_y}$ & $\frac{2\pi}{L_x}$ & $1$ & $10000$             \\[2mm]
           $\Omega$  &   $0.1$   &  $0.5$ & $1.1$ & $-0.1$ &    $\frac{2\pi}{L_y}$ &$\frac{2\pi}{L_x}$ &  $1$ & $10000$             \\[2mm]
           $\vpare$  &   $1$   &  $0$ & $1$ & $0.2$      &    $\frac{2\pi}{L_y}$ & $\frac{2\pi}{L_x}$ & $1$ & $10000$             \\[2mm]
           $\vpari$  &  $1$    &   $0.3$ & $0.9$ & $-0.2$ &     $\frac{2\pi}{L_y}$ & $\frac{2\pi}{L_x}$ & $1$ & $10000$             \\[2mm]
           $t_e$  &   $1$   &  $1.2$ & $0.5$   & $0.1$ & $\frac{2\pi}{L_y}$&  $\frac{2\pi}{L_x}$ &  $1$ & $10000$             \\[2mm]            
           $t_i$  & $1$     &  $1.1$ & $0.4$ & $0.4 $& $\frac{2\pi}{L_y}$ & $\frac{2\pi}{L_x}$ & $1$ &  $10000$             \\[2mm]  
           $\phi$ & $1$ & $3.1$ & $0.8$ & $0.1$ & $\frac{2\pi}{L_y}$ &$\frac{2\pi}{L_x}$ &  $1$ & $10000$             \\[2mm]  
             \hline
    \end{tabular}
    \caption{Parameters imposed for the manufactured solutions $\boldsymbol{\xi}_m$ in dimentionless units. }
    \label{tab:verif_param}
\end{table}
The temporal frequency $A_t$ of the manufactured solutions is set higher than the spatial frequencies in order to ensure that the error is dominated by the temporal discretization, thereby enabling a clear assessment of the time convergence of the method.

The parameters used for the scaling test in Sec.~\ref{sec:parall} and for the convergence test in Tab.~\ref{tab:rde} are: $\rho^{-1}_* = 250$, $\eta_{0e} = \eta_{0i} = 5$, $m_i/m_e = 200$, and $\nu = 0.1$. The numerical tests are performed on a computational domain of size $L_x = 300\rho_{s,0}$ in the radial direction and $L_y = 400\rho_{s,0}$ in the vertical direction. The coefficients $\Tilde{\chi}^e_{\parallel 0}$ and $\Tilde{\chi}^i_{\parallel 0}$ in the flux limiter for parallel conduction are both set to $10$. The diffusion coefficients in the perpendicular direction for $n$, $T_e$, $T_i$, $\Omega$, $\vpare$, and $\vpari$ are $40$.

The two sets of simulations performed for the comparison of the WCT with explicit time stepping in Sec.~\ref{sec:efficiency} are carried out with the following parameters. In the first set, $m_i/m_e=200$, $\nu=0.1$, and the diffusion coefficients in the perpendicular direction for $n$, $T_e$, $T_i$, $\Omega$, $\vpare$, and $\vpari$ are set to $40$. In the second set, $m_i/m_e=1200$, $\nu=0.05$, and the corresponding diffusion coefficients are set to $20$. The remaining parameters are the same for both sets of simulations: $\rho^{-1}_* = 250$ and $\eta_{0e} = \eta_{0i} = 5$. The numerical tests are performed on a computational domain of size $L_x = 300\rho_{s,0}$ in the radial direction and $L_y= 400\rho_{s,0}$ in the vertical direction. The coefficients $\Tilde{\chi}^e_{\parallel 0}$ and $\Tilde{\chi}^i_{\parallel 0}$ in the flux limiter for parallel conduction are both set to $10$.

For the tests conducted with the experimental equilibrium, whose results are
reported in Sec.~\ref{sec:exp}, the parameters used are $\rho^{-1}_* = 450$, $\eta_{0e} = \eta_{0i} = 0.5$, $m_i/m_e = 1200$, and $\nu = 0.1$. The numerical tests are performed on a computational domain of size $L_x = 266\,\rho_{s,0}$ in the radial direction and $L_y = 640\,\rho_{s,0}$ in the vertical direction. The perpendicular diffusion coefficients are set to $40$ for $T_e$, $T_i$ and
$\vpari$, and to $60$ for $n$, $\Omega$ and $\vpare$.

\end{document}